%% file: main.tex
\documentclass[10pt,twocolumn,letterpaper]{article}

\usepackage[pagenumbers]{iccv}

\input{preamble}

\definecolor{iccvblue}{rgb}{0.21,0.49,0.74}
\usepackage[pagebackref,breaklinks,colorlinks,allcolors=iccvblue]{hyperref}

\title{MAMBO: High-resolution Generative Approach for Mammography Images}

\author{
Milica Škipina $^{1,2,3}$ \quad
Nikola Jovišić$^{1,3}$ \quad
Nicola Dall'Asen$^{2,4}$ \\
Vanja Švenda $^{1}$ \quad
Anil Osman Tur $^{2,6}$ \quad
Slobodan Ilic $^{1}$ \\
Elisa Ricci $^{2,5}$ \quad
Dubravko Culibrk $^{1,3}$ \\
\\
$^1$ The Institute for Artificial Intelligence Research and Development of Serbia \quad
$^2$ University of Trento \\
$^3$ Faculty of Technical Sciences, University of Novi Sad \quad
$^4$ University of Pisa \\
$^5$ Fondazione Bruno Kessler \quad
$^6$ University of Verona \\
{\tt\small \{milica.skipina, nikola.jovisic, vanja.svenda, slobodan.ilic, dubravko.culibrk\}@ivi.ac.rs}\\
{\tt\small \{nicola.dallasen, anilosman.tur, e.ricci\}@unitn.it}\\
}

\begin{document}

\maketitle
\input{sec/0_abstract}    
\input{sec/1_introduction}
\input{sec/2_related_work}
\input{sec/3_method}
\input{sec/4_experiments}

\input{sec/5_conclusion}

\section*{Acknowledgments}
This work was supported by the  Junior Fellows Program of EU H2020 project AI4Media (No. 951911) and by the Institute for AI R\&D of Serbia, under the Seed Research Grant Program for young scientists, ("Diffusion Models for Mammography Images Generation and Anomaly Detection"), through the financial support of the Serbia Accelerating Innovation and Entrepreneurship (SAIGE) Project, a joint investment by the Republic of Serbia, Ministry of Science, Technological Development and Innovation, the World Bank and the European Union. Disclaimer: The statements, opinions and data contained in this publication are solely those of the individual author(s) and contributor(s) and not of the SAIGE Project. The SAIGE Project disclaims responsibility for any use that may be made of the information contained therein.
{
   \small
    \bibliographystyle{ieeenat_fullname}
    \bibliography{main}
}
\input{sec/X_suppl}

\end{document}

%% file: preamble.tex
\usepackage{multirow}
\usepackage{graphicx}
\usepackage{mathtools,amssymb}
\usepackage{amsfonts}     
\usepackage{nicefrac}       
\usepackage{microtype}     
\usepackage{pgfplots,pgfplotstable}
\usepackage{array,colortbl}
\usepackage{xcolor}
\usepackage{algorithm,algorithmicx,algpseudocode}
\usepackage{caption}
\usepackage{graphbox}
\usepackage{placeins}
\usepackage{wrapfig}
\usepackage{subcaption}
\usepackage{etoolbox}
\pgfplotsset{compat=1.18}
\usepackage{siunitx}

\newcommand{\patchclean}{\mathbf{x}_{P}}
\newcommand{\gctxclean}{\mathbf{x}_{G}}
\newcommand{\lctxclean}{\mathbf{x}_{L}}
\newcommand{\shiftgctxclean}{\overrightarrow{\mathbf{x}_{G}}}

\newcommand{\gctxnoise}{\mathbf{x}_{T_G}}
\newcommand{\gctx}{\mathbf{x}_{0_G}}
\newcommand{\shiftgctx}{\overrightarrow{\mathbf{x}_{0_G}}}
\newcommand{\shiftgctxs}{\overrightarrow{\mathbf{X}_{0_G}}}

\newcommand{\lctxnoise}{\mathbf{x}_{T_L}}
\newcommand{\lctx}{\mathbf{x}_{0_L}}
\newcommand{\lctxs}{\mathbf{X}_{0_L}}

\newcommand{\pnoise}{\mathbf{x}_{T_P}}
\newcommand{\patch}{\mathbf{x}_{0_P}}
\newcommand{\patches}{\mathbf{X}_{0_P}}
\newcommand{\patchesclean}{\mathbf{X}_{P}}

\newcommand{\netone}{\epsilon_{\theta_1}}
\newcommand{\nettwo}{\epsilon_{\theta_2}}
\newcommand{\netthree}{\epsilon_{\theta_3}}

%% file: sec/0_abstract.tex
\begin{abstract}
Mammography is the gold standard for the detection and diagnosis of breast cancer. This procedure can be significantly enhanced with Artificial Intelligence (AI)-based software, which assists radiologists in identifying abnormalities. However, training AI systems requires large and diverse datasets, which are often difficult to obtain due to privacy and ethical constraints. To address this issue, the paper introduces \textbf{MAM}mography ensem\textbf{B}le m\textbf{O}del (\textbf{MAMBO}), a novel patch-based diffusion approach designed to generate full-resolution mammograms. Diffusion models have shown breakthrough results in realistic image generation, yet few studies have focused on mammograms, and none have successfully generated high-resolution outputs required to capture fine-grained features of small lesions. 
To achieve this, MAMBO integrates separate diffusion models to capture both local and global (image-level) contexts. The contextual information is then fed into the final model, significantly aiding the noise removal process. This design enables MAMBO to generate highly realistic mammograms of up to 3840×3840 pixels. Importantly, this approach can be used to enhance the training of classification models and extended to anomaly segmentation. Experiments, both numerical and radiologist validation, assess MAMBO's capabilities in image generation, super-resolution, and anomaly segmentation, highlighting its potential to enhance mammography analysis for more accurate diagnoses and earlier lesion detection. The source code used in this study is publicly available at: \url{https://github.com/iai-rs/mambo}.
\end{abstract}

%% file: sec/1_introduction.tex
\section{Introduction}
\label{sec:intro}

\begin{figure}[htpb]
    \centering
    \includegraphics[width=\columnwidth]{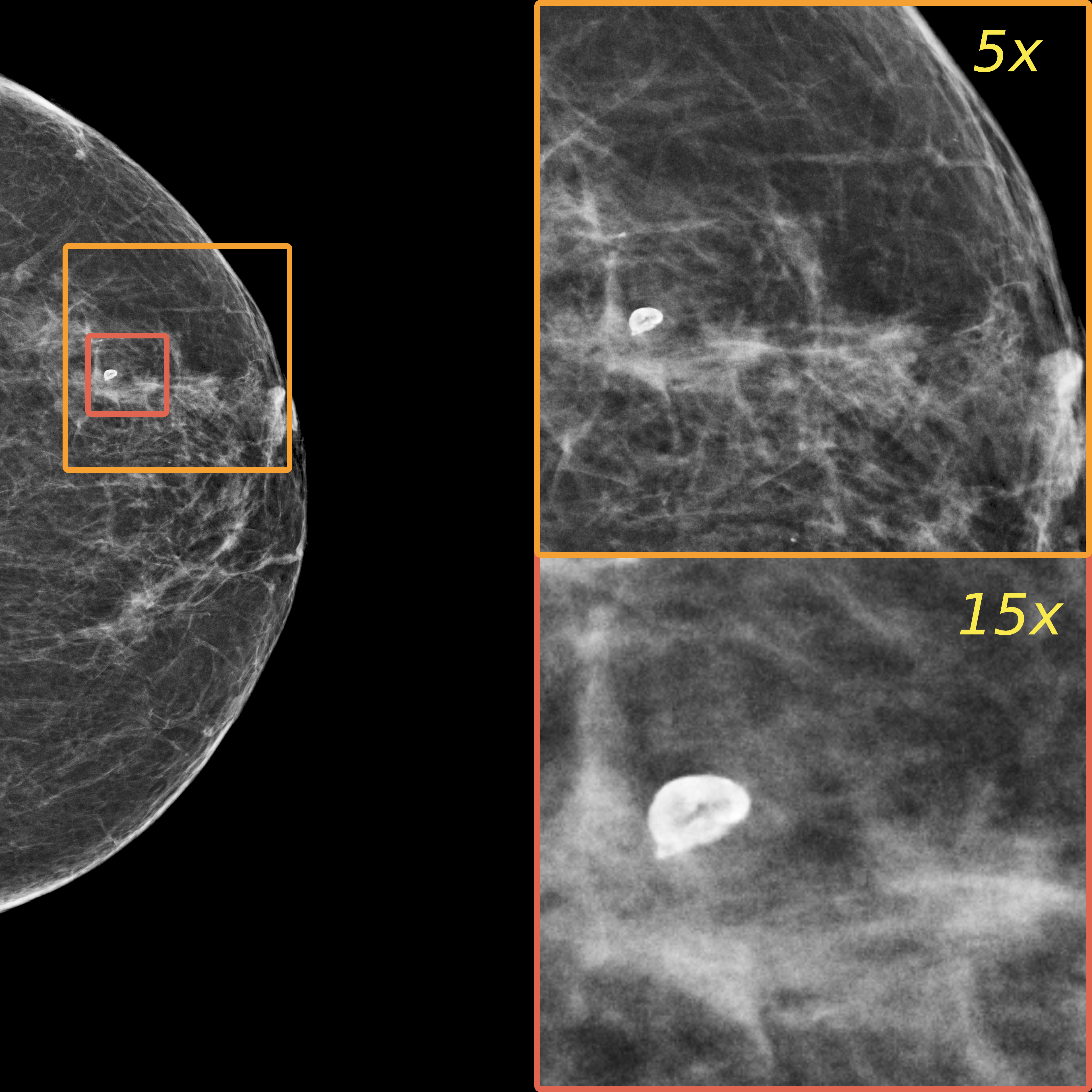}
    \caption{Synthetic $3840x3840$ mammogram generated using MAMBO. Details at different resolutions correspond to the global context (whole image),  \textcolor[rgb]{1,0.6,0}{local context}, and individual \textcolor[rgb]{1,0.4,0.4}{patch}. Best viewed when zoomed in.}
    \label{fig:teaser-combined}
\end{figure}

Breast cancer is the most common malignancy and leading cause of cancer deaths in women around the world \cite{incidence}. 
Mammography is the preferred method for the detection and prevention of breast cancer, as it is effective, cost-effective, and reliable, making it ideal for screening large populations. This, in turn, produces numerous mammograms, all of which should be analyzed by qualified medical professionals.
Analysis often requires considerable effort, as one needs to detect subtle contrast variations caused by differences in breast density \cite{acr}.
To address this challenge, radiologists have increasingly relied on computer-aided diagnosis (CAD) systems, which have become essential tools in mammography in screening centers and hospitals \cite{doi2007computer}.  
Advancements in Deep Learning (DL) have brought further improvements, setting a new standard in medical imaging, and enhancing both diagnosis accuracy and patient care \cite{mall2023comprehensive}. 
Provided that sufficient diverse training data is available, a wide range of tasks can be tackled with DL \cite{ai-in-mamo}: image generation \cite{monai}, risk prediction \cite{yala}, cancer presence prediction \cite{e2e}, detection \cite{yolo}, segmentation \cite{segmentation}, and anomaly detection \cite{anomaly}. 

Among clinical imaging modalities, mammography provides the highest spatial resolution \cite{rangarajan2022ultra} and full-field digital mammography (FFDM) systems usually produce images at very high digital resolutions, ranging from $1920\times2304$ to $4708\times5844$ pixels, where pixel size corresponds to an area of \SI{100}{\micro\metre} and \SI{50}{\micro\metre}, respectively \cite{keavey2012comparison, liu2014evaluation}. This high resolution is necessary to observe microcalcifications and cancers less than 1 cm in size, which has a significant impact on the outcome of the therapy. The 10-year survival rate drops from more 95\% in patients in whom the observed cancer is less than 1 cm to about 60\% when the cancer size is more than 3 cm~\cite{zheng2015effect}.

However, obtaining high-quality medical images can be challenging, often due to privacy and legal concerns. Thus, there is growing interest in DL-driven approaches for generating synthetic data \cite{synth}, which can be thought of as a form of advanced data augmentation. One such approach are the denoising diffusion models~\cite{diffusion}, which have recently been adapted to radiology ~\cite{chest}. However, they typically support relatively low image resolutions (e.g., $ 256\times 256$ pixels), limiting their usefulness in mammography applications.   

To overcome these challenges, this paper introduces \textbf{MAM}mography ensem\textbf{B}le m\textbf{O}del (\textbf{MAMBO}), a novel approach that leverages an ensemble of diffusion models to generate mammography images at their native resolution. MAMBO uses a patch-based approach for generation which is conditioned both on local and global (full-image) context. In a nutshell, MAMBO involves the development of three distinct models: the first generates ``standard''-resolution ($256\times256$ pixels) images to provide global context; the second increases the resolution to create local context for the target patch; and the third combines the outputs of both models to guide the generation of high-resolution patches, which are then reconstructed into a full-resolution synthetic mammogram. An example of a synthetic image generated by MAMBO is shown in ~\cref{fig:teaser-combined}. The image presents details at multiple resolutions, corresponding to the global and local contexts, and individual patch at full resolution. To the best of our knowledge, this is the first work to propose a diffusion model approach for generating very high-resolution synthetic mammograms.

The proposed approach is evaluated  on various tasks using three datasets: VinDr \cite{vindr}, RSNA \cite{rsna}, and InBreast \cite{moreira2012inbreast}. 
The evaluation demonstrates the effectiveness of MAMBO in generating realistic synthetic mammograms, using standard generative model metrics (e.g., Fr\'echet Inception Distance - FID) and comparisons with prior methods. Expert radiologists were unable to reliably distinguish synthetic images from real mammograms in blinded evaluations. A classification model trained with a dataset composed of 50\% synthetic mammography images generated by MAMBO demonstrates improved performance, increasing recall by 35\% and F1-score by 14\% compared to a model trained solely on real data. In addition, we present the results for anomaly segmentation as a downstream task. This is achieved by generating a healthy counterpart for an input mammogram and comparing it to the original image. The approach makes pixel-wise predictions on the presence of a lesion without training on pixel-wise labels. Considering the scarcity of labeled medical data, such an unsupervised approach is very valuable. Results as high as 0.423 are achieved in Intersection over Union (IoU). 

%% file: sec/2_related_work.tex
\section{Related Work}
\label{sec:related}

\subsection{Synthetic Image Generation in the Medical Domain}
When it comes to generating mammography images, whether to train other models or as a foundation for anomaly detection, Generative Adversarial Networks (GANs) remain the core technology. Oyelade \textit{et al.} \cite{gan-region} provided a comprehensive overview of the different approaches employed in this domain. The paper also proposed an approach for synthesizing specific regions of interest in mammography. While the approach can be considered patch-based, it can not generate whole mammograms. Park \textit{et al.} \cite{gan} focused on generating whole images of healthy breasts and showed reasonable performance in terms of detecting cancer as anomalies \cite{gan-anomaly}. This type of approach, similarly to the one proposed in this study, provides the opportunity to create pixel-level segmentation masks of lesions, using only weak labels in the form of image-level annotations. Using the StyleGAN2 architecture \cite{karras2020analyzing} for generation, they achieved an impressive FID score of 4.383.

Recently, diffusion models have been widely adopted in computer vision \cite{croitoru2023diffusion}, demonstrating impressive generative capabilities. As a result, there is growing research interest in applying these models to the field of radiology. For example, it has been shown to be effective for weakly-supervised anomaly detection \cite{anomaly}, particularly when modifications to the diffusion process itself are made \cite{wyatt2022anoddpm}.

A study by Pan \textit{et al.} showed that diffusion models are able to generate X-ray, MRI and CT images that are indistinguishable from the originals even for the eye of the professional radiologist \cite{chest} and therefore pass the visual Turing test.  
The study, however, did not use any mammography data, and the resulting images are of relatively low resolution ($256\times256$ pixels). 

When it comes to generative AI in medicine, one should mention the efforts directed towards a standardized baseline framework for it \cite{monai}. The models available, however, are not capable of generating high-resolution mammograms yet.

\subsection{High-Resolution Image Generation}
For high-resolution images, using the Latent Diffusion Model (LDM) is common \cite{rombach2022high}. This approach first compresses the image, typically using a trained Variational Autoencoder (VAE) \cite{kingma2013auto} or a Vector-Quantized Variational Autoencoder (VQ-VAE) \cite{van2017neural}, thus transferring the image to a latent representation. Noising and denoising are applied to this latent representation, which is then transformed back into the pixel space. This approach helps to reduce the computational cost, which would be too high if the diffusion model were trained directly on high-resolution images.

An alternative to LDM for high-resolution images is the patch-based approach, which is used in this study. While we are unaware of any applications of this method in mammography, Ding \etal recently proposed a technique called Patch-DM \cite{neighbour-patch} for natural images. They use a sliding window approach, with the local context provided by padding the patch and passing the overlap of the padding with the neighboring patches in the latent space. Features extracted from a pretrained CLIP model \cite{radford2021learning} are leveraged as global conditions, along with classifier-free guidance \cite{ho2022classifier}, to enhance training speed and stability. 

\subsection{Knowledge-based In Silico Models}
In addition to deep generative models, knowledge-based methods offer an alternative approach for generating synthetic mammograms \cite{badano2018evaluation, sizikova2023knowledge}. These techniques combine digital breast phantoms with physics-based X-ray simulation, providing explicit control over anatomical structures, lesion placement, radiation dose, and imaging parameters.

While these pipelines support controlled experiments with ground-truth labels, the diversity of generated images is constrained by the parameter space of the anatomical and pathological models, as well as the fixed rules of the acquisition simulation. As qualitatives in \cite{sizikova2023knowledge} show, this can lead to repetitive textures, simplified lesion morphology, and reduced visual realism. In contrast, diffusion models can overcome these limitations by learning rich, high-resolution anatomical structures directly from data and generate diverse, realistic images using identical conditioning inputs.

%% file: sec/3_method.tex
\begin{figure*}[htp]
\centering
\includegraphics[width=\textwidth]{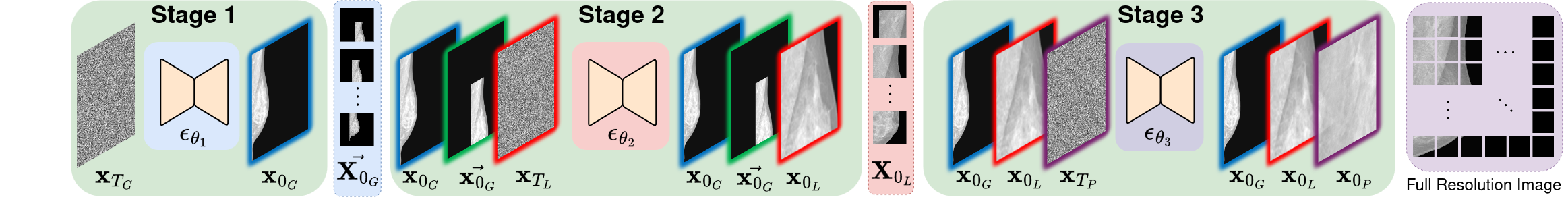}
\caption{
In the three-stage approach of MAMBO, the first stage of the model generates a novel \textcolor{RoyalBlue}{global context} $\gctx$, which is then used to generate a set of \textcolor{OrangeRed}{local contexts} $\lctxs$ in the second stage, conditioned on \textcolor{ForestGreen}{shifted global contexts} $\shiftgctxs$. Synthetic global context and synthetic local contexts become the conditioning in the third stage to generate \textcolor{Plum}{highly detailed patches} $\patches$, which are finally combined to obtain a high-resolution full mammogram.
}
\label{fig:method}
\end{figure*}

\section{Method}
\label{sec:method}

We propose a novel approach to generate high-resolution synthetic mammography images using an ensemble of three diffusion models in a coarse to fine manner. This allows the number of parameters in each model to be kept roughly constant and at a reasonable amount for consumer-level hardware, while creating full-resolution results. 
 
Once all the patches of a full-resolution mammogram are generated, they are overlapped to generate output images without visible artifacts.
The pipeline consists of three Denoising Diffusion Probabilistic Models (DDPM), each operating on $s\times s$ pixel images but with progressively more details, starting from the whole image, then moving to a patch corresponding to a resized sub-region of the original image, and, finally, full-resolution $s\times s$ patches of the original image.

The low-resolution model generates new synthetic images using a single-channel U-Net, yielding a low-resolution version of the complete mammogram. For the subsequent two models, a patch-based approach is employed. As mammography images are grayscale, the standard 3-channel U-Net is modified to accommodate this data format. Only one of the input channels is used to represent mammogram patches, while the additional channels are used to provide the global and local context of the patch. This allows incorporating sufficient information to generate patches with enough details and, at the same time, to maintain consistency between neighboring patches.

\subsection{Denoising Diffusion Probabilistic Model}
Diffusion models are a class of generative models that transform simple noise into a complex data distribution through a gradual denoising process. The core principle behind DDPM \cite{diffusion} is to learn to reverse the Markov process that starts with the data distribution and applies a series of transformations to reach the noise distribution. This process is parameterized by a neural network that predicts the noise to be removed at each step, gradually reconstructing the data from pure noise.

The forward diffusion process is defined as:
\begin{equation}
q(\mathbf{x}_{t}|\mathbf{x}_{t-1}) = \mathcal{N}(\mathbf{x}_{t}; \sqrt{1-\beta_t}\mathbf{x}_{t-1}, \beta_t\mathbf{I}),
\end{equation}
where $\mathbf{x}_{t}$ represents the data at a certain timestep $t$, $\beta_t$ is a variance schedule, and $\mathcal{N}$ denotes the Gaussian distribution. This process incrementally adds Gaussian noise over $T$ timesteps, resulting in $\mathbf{x}_{T}$ being isotropic Gaussian noise.

The reverse process is modeled as follows:
\begin{equation}
p_{\theta}(\mathbf{x}_{t-1}|\mathbf{x}_{t}) = \mathcal{N}(\mathbf{x}_{t-1}; \mu_{\theta}(\mathbf{x}_{t}, t), \Sigma_{\theta}(\mathbf{x}_{t}, t)),
\end{equation}
where $\mu_{\theta}(\mathbf{x}_{t}, t)$ and $\Sigma_{\theta}(\mathbf{x}_{t}, t)$ are learned functions that predict the mean and variance of the reverse Gaussian transition. The objective of training diffusion models is to optimize the parameters $\theta$ of these functions to accurately reconstruct the original data distribution from the noise.  

\subsection{An Ensemble of Diffusion Models}

As mammography images are grayscale, a straightforward baseline is to apply a single channel U-Net to the full resolution $s \times s$ pixel patches extracted from the mammograms. This approach, without any additional information, cannot create a complete mammogram (see \cref{sec:qualitatives}).

Clearly, context needs to be provided, and, arguably, the simplest way to do so is to provide context in the form of an additional input channel, as it requires fewer architectural modifications. As the aim is to generate full-resolution mammograms, 

the model must be able to generate a consistent grid of patches. Therefore, just providing context in the form of a low-res representation of the whole image is not sufficient. Rather, the notion of ``locality'' of the patch is needed, and while this could be achieved in several ways, we provide an additional channel to the network, as this allows including additional guidance information in the future. 

This leads to a patch-based DDPM model based on the standard, three-channel U-Net architecture, within which the context information is provided at two scales (global and local) using the additional input channels, as shown in \cref{fig:method}: 

\begin{equation}
    \label{eq:patch_generation}
    \patch = \netthree(\pnoise \vert \gctxclean, \lctxclean), \pnoise \sim \mathcal{N}(0, \mathbf{I}),
\end{equation}

where $\patch$ represents the generated patch, and $\gctxclean$ and $\lctxclean$ represent the global and local context, respectively. This model can be readily used to create high-resolution mammograms that are visually indistinguishable from the original.

To generate a completely synthetic image, the global and local context must also be generated, and to this end, two additional diffusion models are trained. A single-channel U-Net $\netone$ is used to generate the global context at low resolution starting from pure noise $\gctxnoise$:

\begin{equation}
    \gctx = \netone(\gctxnoise), \gctxnoise \sim \mathcal{N}(0, \mathbf{I}),
\end{equation}

$\gctx$ then replaces $\gctxclean$ as the global context in \cref{eq:patch_generation}.

Conversely, we use a three-channel U-Net to generate a downsized local context of the full-resolution mammogram:

\begin{equation}
    \lctx = \nettwo(\lctxnoise \vert \gctx, \shiftgctx), \lctxnoise \sim \mathcal{N}(0, \mathbf{I}),
\end{equation}

$\lctx$ then replaces $\lctxclean$ as the local context in \cref{eq:patch_generation}. $\shiftgctxclean$ is obtained by shifting the original image and gives a notion of locality for the local context generation.

For efficiency, once generated, local context patches are merged to create mid-resolution images, effectively upscaling the low-resolution (global-context) image. 

Using a similar approach, the final, high-resolution patch generator adds even more details and upscales images an additional $k$ times, with $k$ being an hyperparameter.

\subsubsection{Training}
\label{section:training}

The final (high-resolution) model generates $s\times s$-pixel patches $\patches$ of a full-resolution mammogram. To create training data, patches $\patchesclean$ of size $s\times s$ are first randomly cropped from the original image. These patches represent the first channel of the input. Then, the local context $\lctxclean$ is obtained by cropping a $k$ times larger part of the original. That is, the local context of size $k\times k$ is cropped, where the patch in the previous step represents the central square. Finally, the third channel of the input is the whole image $\gctxclean$. 

To combine data along the channel dimension, local and global context are resized to have spatial dimensions $s \times s$.

A similar procedure is used to generate training data for the mid-resolution (local context) model, where the first channel now corresponds to $(s \times k) \times (s \times k)$-pixel patches of the original images, scaled down to $s\times s$ pixels, and the second channel $\shiftgctxclean$ is obtained by shifting the whole image to have the center of the global context represent the center of the cropped patch. 

The noising procedure differs from the original DDPM approach~\cite{diffusion}. As illustrated in \cref{fig:method}, noise is only added and removed to the first channel in the data, which represents the image patch. The network is fed with the global and local context channels as is, regardless of the timestep $t$, and this procedure is presented in \Cref{alg:training}.  The loss is therefore computed only on the first channel. At inference time, the noise is removed only from the first channel, as shown in \cref{alg:sampling}.

\algrenewcommand\algorithmicindent{0.5em}%

\begin{algorithm}[H]
  \caption{Training with data channel conditioning} \label{alg:training}
  \begin{algorithmic}[1]
    \Repeat
      \State $\patchclean \sim q(\mathbf{x}_0)$
      \State $t \sim \mathrm{Uniform}(\{1, \dotsc, T\})$
      \State $\boldsymbol{\epsilon}\sim\mathcal{N}(\mathbf{0},\mathbf{I})$
      \State $\begin{aligned}[t]
        &\mathbf{x}_t^1 = \sqrt{\bar\alpha_t} \patchclean
        + \sqrt{1-\bar\alpha_t}\boldsymbol{\epsilon} \\
        &\mathbf{x}_t^{2,3} = \lctxclean, \gctxclean 
      \end{aligned}$
      \State Take gradient descent step on the whole
      \Statex $\qquad \nabla_\theta \left\| \boldsymbol{\epsilon} - \netthree(\mathbf{x}_t, t) \right\|^2$
    \Until{converged}
  \end{algorithmic}
\end{algorithm}

\begin{algorithm}[H]
  \caption{Sampling with data channel conditioning} \label{alg:sampling}
  \begin{algorithmic}[1]
    \vspace{.04in}
    \State $\begin{aligned}[t]
        &\mathbf{x}_T^1 = \pnoise \sim \mathcal{N}(\mathbf{0}, \mathbf{I}) \\
        &\mathbf{x}_T^{2,3} = \lctxclean, \gctxclean \textbf{ is previously generated}
    \end{aligned}$
    \For{$t=T, \dotsc, 1$}
      \State $\mathbf{z} \sim \mathcal{N}(\mathbf{0}, \mathbf{I})$ if $t > 1$, else $\mathbf{z} = \mathbf{0}$
      \State $\begin{aligned}[t]
      &\mathbf{x}_{t-1}^1 = \frac{1}{\sqrt{\alpha_t}}\left(\mathbf{x}_t^1 - \frac{1-\alpha_t}{\sqrt{1-\bar\alpha_t}} \netthree(\mathbf{x}_t, t) \right) + \sigma_t \mathbf{z} \\
      &\mathbf{x}_{t-1}^{2,3} = \mathbf{x}_T^{2,3}
      \end{aligned}$
    \EndFor
    \State \textbf{return} $\mathbf{x}_0$
    \vspace{.04in}
  \end{algorithmic}
\end{algorithm}

\subsubsection{Patch Aggregation and Overlap}
\label{section:inference}

Although local context helps maintain the continuity between the patches, generating images without patch level overlap causes a visible checkerboard effect and additional discontinuity artifacts. 

Thus, only the first patch of the image is generated independently, and for all subsequent patches, the overlapping region is taken into account by transferring it from already generated neighboring patches. During the denoising process, at each timestamp $t$, the corresponding amount of noise is added to the overlapping part. This ensures a smooth transition between the newly generated data and their pre-existing neighbors. 

\subsection{Anomaly Segmentation}
 Generative models are often used within an anomaly segmentation pipeline \cite{xia2022gan, wolleb2022diffusion, bercea2024diffusion}. This enables the model to make pixel-wise predictions on lesion presence, while not requiring pixel-wise annotations for training. Therefore, the proposed models' effectiveness is evaluated for this downstream task. To do so, the whole image model is used, trained only on healthy images without any lesions, therefore capturing the distribution of images of breasts that consist entirely of healthy tissue. A limited amount of Gaussian noise is added to the diseased images, which are then denoised using the proposed model. As it is capable of generating only healthy tissue, the lesions from the image are not reconstructed, and the surrounding healthy tissue is well-replicated. By comparing the original mammogram with the denoised version, the map of anomalies present in the image is obtained.

Noise is added for $\lambda$ steps, using the forward noising procedure used in DDPM, where \(0 < \lambda < T\).  $\lambda$ is a hyperparameter that determines the start point of denoising, balancing original information preservation (lower $\lambda$) and anomaly removal (higher $\lambda$). In the latter case, the difference of non-anomalous parts might be too big, and even the breast shape and tissue density might change. 

To automatically compare the most important regions of the image, a binary segmentation mask is first needed that precisely outlines the breast area and distinguishes it from the surrounding background.

We can use the breast mask derived from the original image to extract the region of interest in both the original and the denoised image. We then perform histogram matching on these two images, as this accounts for all the changes that the diffusion process introduces in the overall pixel intensities, which could adversely influence the anomaly segmentation process and create spurious differences.

Once this is done, the initial anomaly map is created by subtracting the denoised image from the original. As the breast cancer, including masses, appears white in mammograms \cite{pinsky2010mammographic}, this anomaly map is expected to contain more positive than negative values in the anomalous regions on average. Therefore, all negative values are set to zero. This could analogously be used in detecting dark lesions (\eg some types of cysts) by inverting the values in the image.

Finally, changes that are less than $30\%$ of the intensity of the pixel with the highest intensity are discarded. This step is performed to remove all the non-significant high-frequency changes introduced by the diffusion process. On top of this, a Gaussian blur is applied to further smooth the high-frequency noise and get the final anomaly map.

%% file: sec/4_experiments.tex
\section{Experiments and Results}
\label{sec:results}

MAMBO is evaluated in four application scenarios, highlighting its versatility: mammogram generation, super-resolution, classification and anomaly segmentation. 

The experiments are carried out on open datasets, VinDr \cite{vindr}, RSNA \cite{rsna}, and InBreast \cite{moreira2012inbreast}. 
For image generation and super-resolution tasks, performance is evaluated using the Fréchet Inception Distance (FID) metric \cite{heusel2017gans} and the Learned Perceptual Image Patch Similarity (LPIPS) metric \cite{zhang2018unreasonable}. To assess the performance of anomaly segmentation, Intersection over Union (IoU) is used. For the classification task, AUC score, recall, precision, and F1 score are evaluated. The input dimension $s$ of the U-Net implementation ~\cite{diffusion} is set to $256$ pixels to balance detail capture with simplicity and ease of training. The number $k$ of neighboring patches in the local context is set to $3$, as justified in ~\cref{sec:ablation}. 

\subsection{Mammogram Generation}

\subsubsection{Baselines} 
To the best of our knowledge, MAMBO is the first AI method to generate mammography images in their native resolution. 
Since no other model is directly applicable in this setting, comparison is done to several state-of-the-art approaches that perform specific parts of its functionality.

The approach of Park \etal ~\cite{gan} is the most relevant as it uses generative models, in particular GANs, to create synthetic mammography images. 
However, images are generated at $512\times512$ pixels, far from full-resolution, and two FID scores of 10.425 and 4.383 are reported, without clarifying the experimental setting. The datasets used contain 105,948 normal mammograms, but are not publicly available. The unavailability of the data and the different resolution make a direct comparison unfeasible.

Pan \etal ~\cite{chest} propose a 2D image synthesis framework for the generation of various radiology images based on a diffusion model using a Swin Transformer \cite{liu2021swin}-based network. 

Each model is trained on images of resolution $256\times256$ pixels. An average FID value of 37.27 is reported, obtained from all datasets, with a minimum FID of 22.33 for chest X-ray images. As mammography benchmarks are not included, direct comparison is unfeasible.

Montoya-del-Angel \etal present MAM-E, a pipeline of models capable of generating synthetic lesions on specific regions of the breast. A Stable Diffusion model \cite{rombach2022high} is fine-tuned using mammography images to generate healthy mammograms controlled by text input. Later, Stable Diffusion was used to inpaint synthetic lesions in desired regions. Although radiologists struggled to identify the differences between real and synthetic images, due to the model architecture, the approach is limited to a resolution of $512\times512$. Also, the authors did not report the FID score, which makes a direct comparison with their approach difficult.

\begin{figure*}[htbp]
    \centering
    \begin{subfigure}[b]{0.19\textwidth}
        \includegraphics[width=\textwidth]{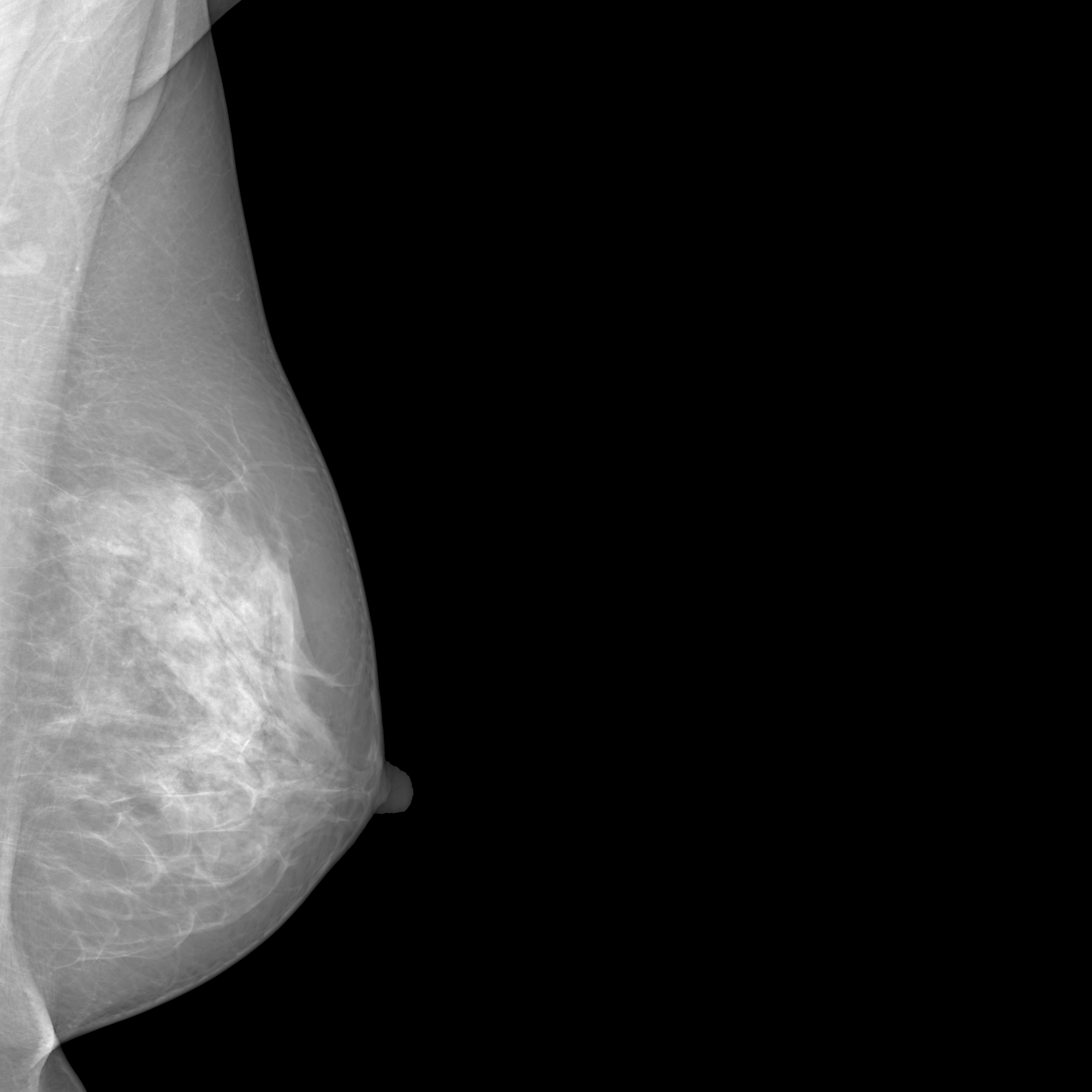}
        \caption{Image from dataset}
        \label{fig:patch_orig}
    \end{subfigure}
    \begin{subfigure}[b]{0.19\textwidth}
        \includegraphics[width=\textwidth]{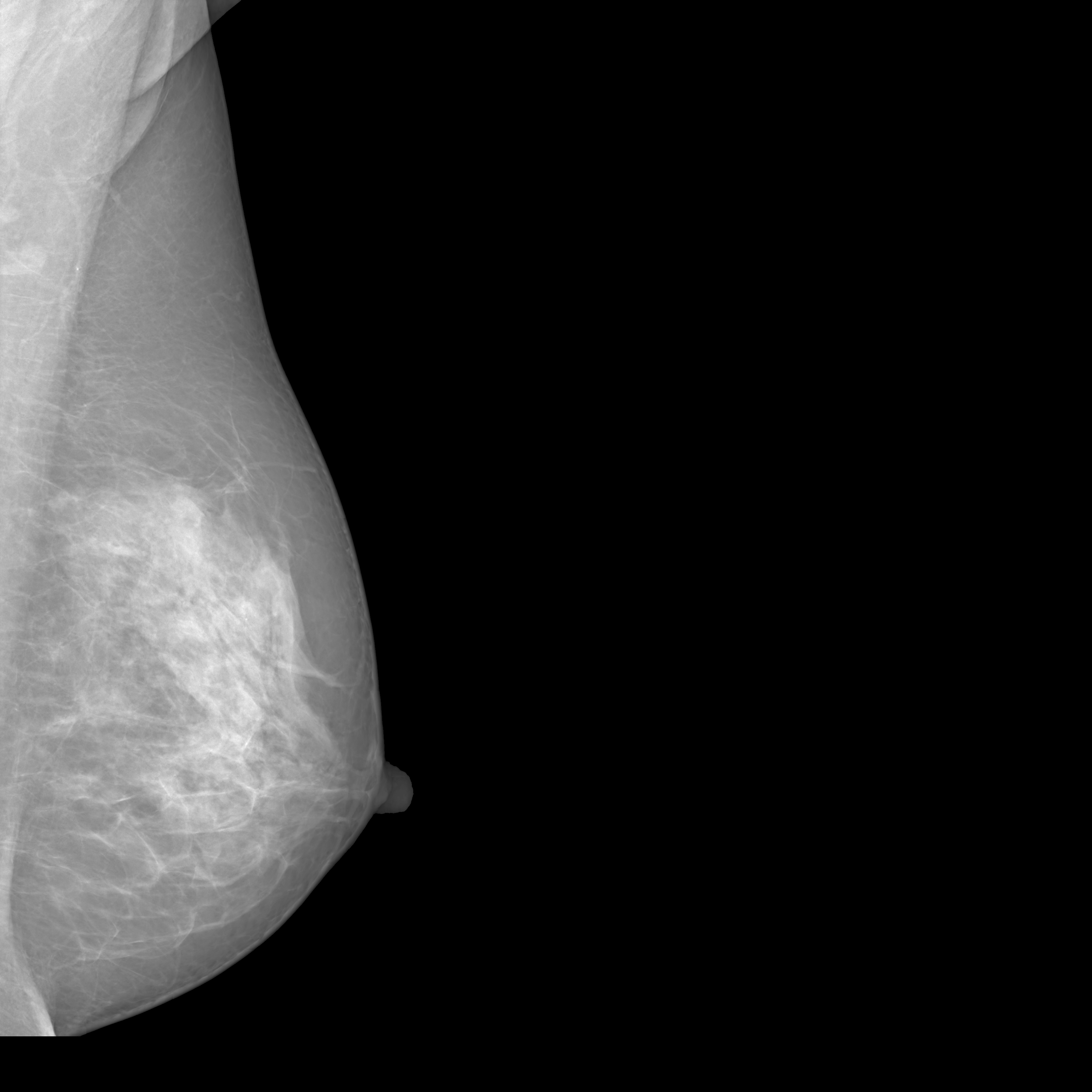}
        \caption{Denoised (OC)}
        \label{fig:patch_3c}
    \end{subfigure}
    \begin{subfigure}[b]{0.19\textwidth}
        \includegraphics[width=\textwidth]{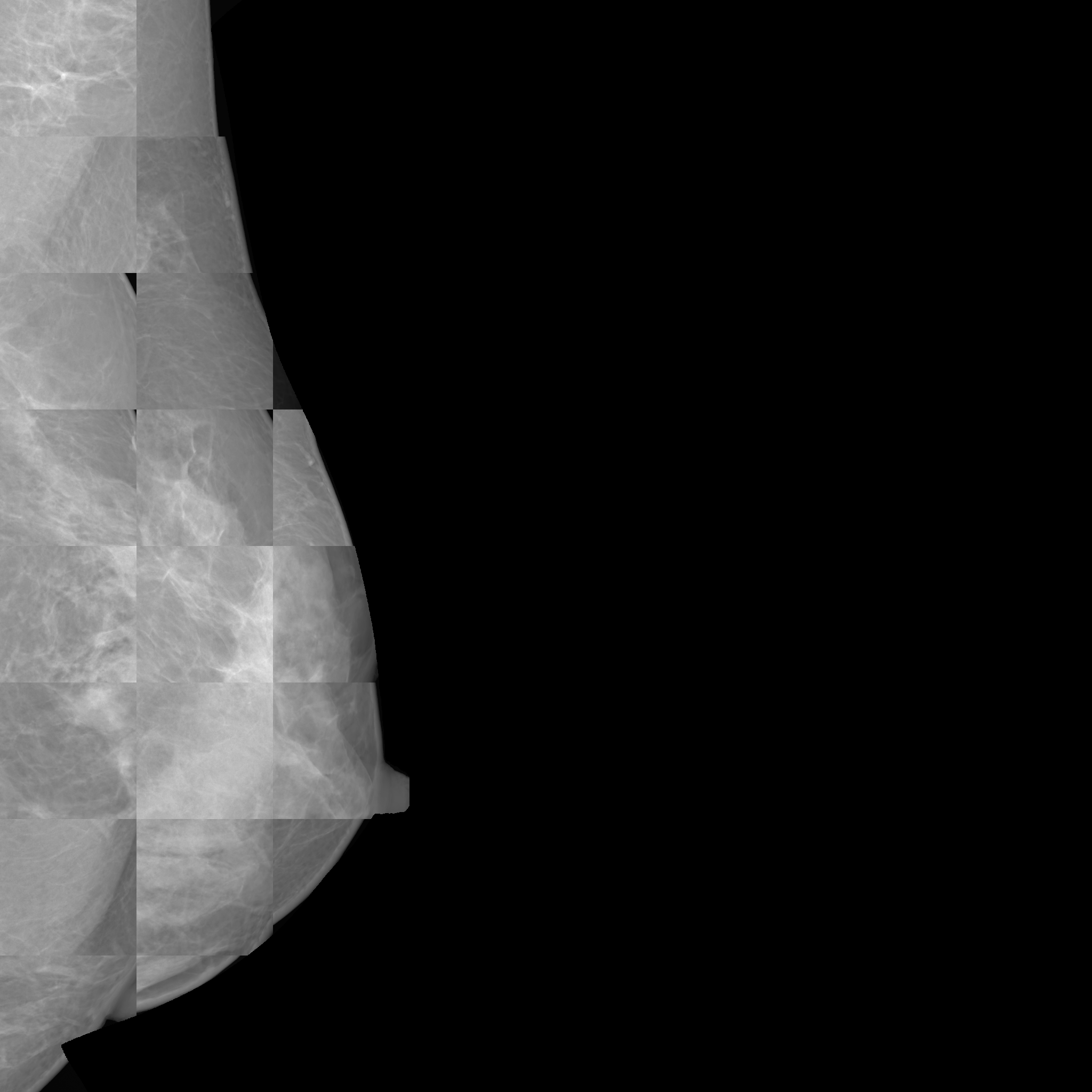}
        \caption{Denoised baseline}
        \label{fig:patch_1c}
    \end{subfigure}
    \begin{subfigure}[b]{0.19\textwidth}
        \includegraphics[width=\textwidth]{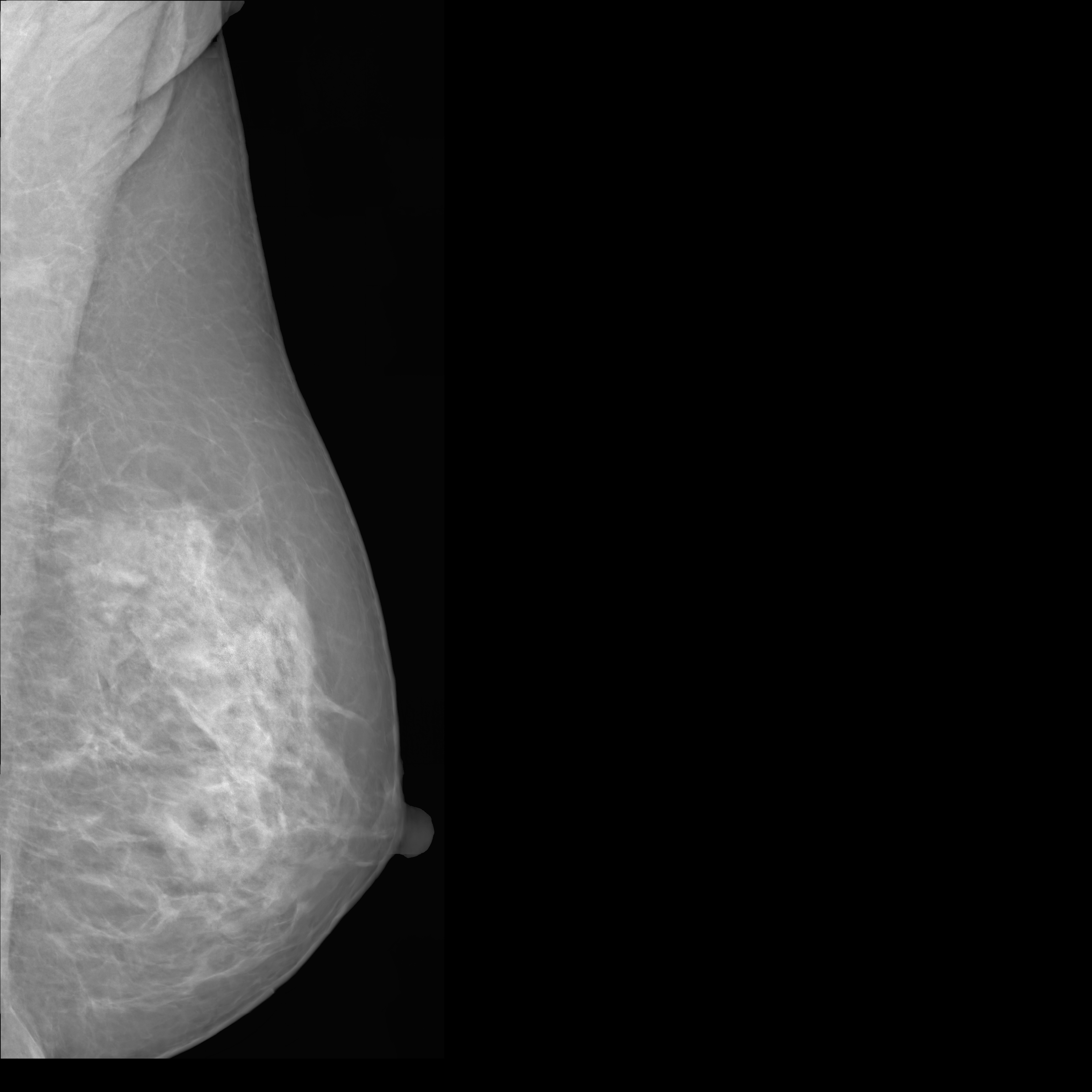}
        \caption{Denoised (GC)}
        \label{fig:patch_pipeline_whole}
    \end{subfigure}
    \begin{subfigure}[b]{0.19\textwidth}
        \includegraphics[width=\textwidth]{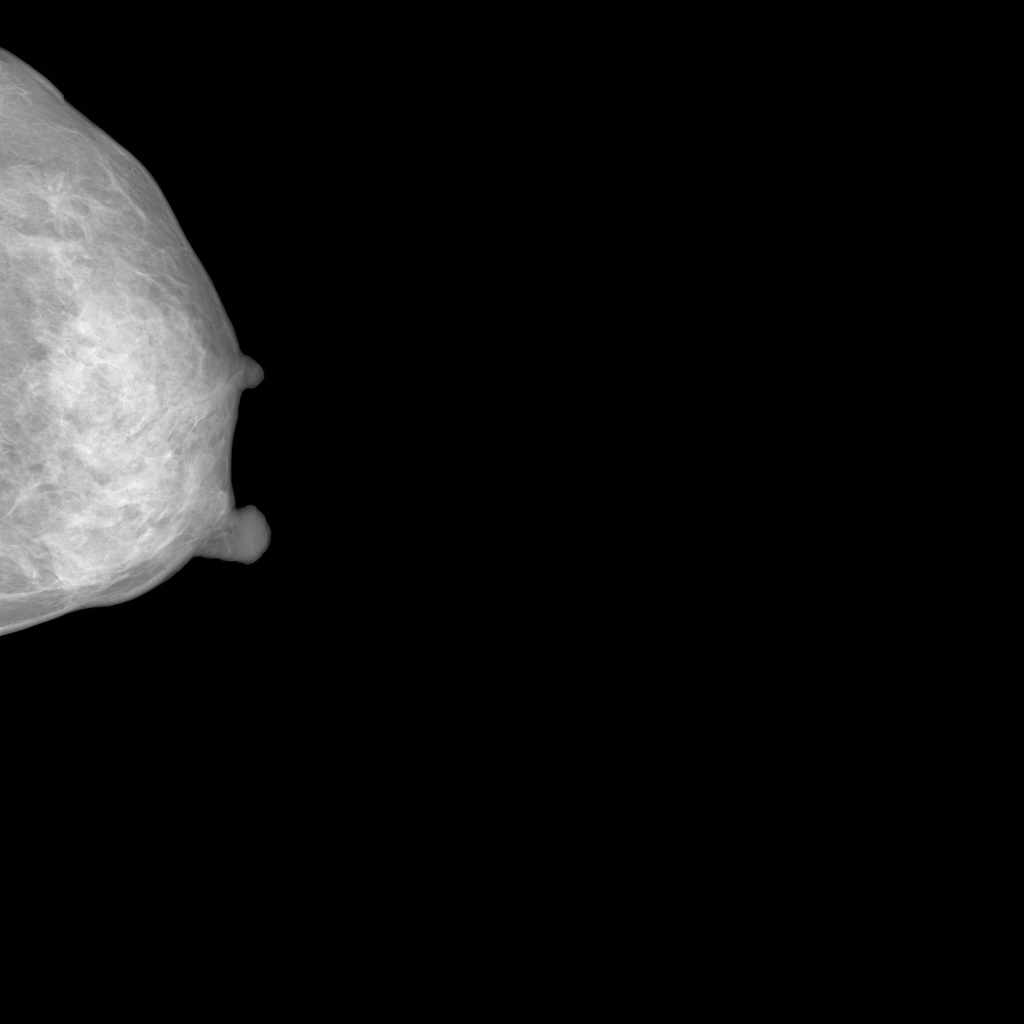}
        \caption{Patch-DM \cite{neighbour-patch}}
        \label{fig:patchdm}
    \end{subfigure}
    \caption{Sample results achieved by MAMBO and PatchDM. (a) Original full-res image. (b) Image denoised from pure noise with the MAMBO pipeline using original local and global contexts from image (a). (c) Image denoised from a partially noisy original with t=750 using a baseline single-channel patch-based model. (d) Image generated with MAMBO Stage 2 and Stage 3 models using the resized original image as a global context. (e) Image generated by Patch-DM in $1024\times1024$ resolution.
     }
    \label{fig:main}
\end{figure*}

The methodology presented in Patch-DM \cite{neighbour-patch} is closest to the proposed approach. It focuses on synthesizing high-resolution natural images using a patch-based approach and a denoising diffusion model. Unlike MAMBO, which works in pixel space, Patch-DM constructs images by independently denoising patches and composing them in feature space using positional embeddings. Pretrained CLIP \cite{radford2021learning} is used to obtain initial global context embeddings, which are later optimized during the training. To explore Patch-DM's potential in the medical domain, we trained it on mammography data to generate $1024\times1024$ images using its original settings with a patch size of $64\times64$ pixels, and we use Mammo-CLIP \cite{ghosh2024mammo} as a pretrained model for obtaining the initial image embeddings. The model was trained for more than two weeks on two NVIDIA A100 GPUs, during which it was exposed to more than 15.5 million mammography images. Despite this extensive training, Patch-DM still struggles to reproduce globally coherent and anatomically realistic mammograms, as shown in \cref{fig:patchdm}. This highlights both the limitations of the model in capturing domain-specific structure and its substantial computational demands.

Sizikova \etal introduced M-SYNTH~\cite{sizikova2023knowledge}, a synthetic digital mammography dataset generated using VICTRE~\cite{badano2018evaluation} breast phantoms and Monte Carlo X-ray simulations across varying tissue densities, lesion sizes, and radiation doses. However, its realism and generalizability are limited by the constrained parameter space of the simulation pipeline. Moreover, because there are no corresponding real images, standard image quality metrics cannot be used to assess image fidelity quantitatively.

\subsubsection{Quantitative Results}

The quantitative performance of each MAMBO part is presented in ~\cref{tab:image_generation}. To evaluate the ability of MAMBO to generate high-quality data when authentic global and local context information is available, four random patches are generated from every image for each dataset, using local and global contexts extracted from the original images. Stratified sampling is used in terms of patch location, so the data set consists of pairs of patches sampled at the same location. One is taken from an original image and one from an image generated from pure noise but using the original local and global context. MAMBO is validated through two experimental setups. 

\input{tables/image_generation}

To evaluate the ability to generate individual patches, the original local and global context was used, and individual patches were generated from noise. This setting is named Region Of Interest (ROI). This resulted in an FID of 2.73 for RSNA and 4.77 for VinDr. To validate whether these results are satisfactory, a lower bound of the FID is computed by dividing the original patches into two subsets and computing the FID between them. This leads to an FID of 0.82 for RSNA and 0.78 for VinDr. In addition, the FID between the random original patches and the random generated patches was computed, leading to a FID of 3.21 and 5.25 for the two data sets, respectively. The same procedure is used to evaluate the performance of local context generation.

In the second setup, the global context model was used to generate $256\times256$-pixel versions of whole mammograms. The same number of synthetic images was generated as in the original dataset. MAMBO showed good performance by achieving an FID of 4.71 for RSNA and 2.14 for VinDr. 

Results for the full MAMBO pipeline, starting from noise to generate global and local context, and then the individual patches, are shown in the \textit{pipeline} column of ~\cref{tab:image_generation}. 

\subsubsection{Qualitative Results}
\label{sec:qualitatives}
The baseline for this task is a single-channel U-Net-based DDPM trained on full-resolution $256\times256$-pixel patches extracted from mammograms. This model learns patch characteristics but performs poorly in image reconstruction tasks. Even if the patch is not generated from pure noise but from the partially noisy patch of an original image, the reconstruction is poor, as shown in \cref{fig:patch_1c}.

What MAMBO can achieve in terms of qualitative results is presented in \cref{fig:teaser-combined} and \cref{fig:patch_pipeline_whole}. 
MAMBO is able to generate high-quality images that are visually indistinguishable from the originals, representing plausible mammogram data to layman eyes. Results are also validated with expert radiologists, with quantitative results shown in \cref{sec:experts}.

~\cref{fig:patch_pipeline_whole} illustrates what MAMBO can achieve in terms of whole mammogram generation. When using global and local context data extracted from an original image, the denoised image (\cref{fig:patch_3c}) is difficult to distinguish from the original (shown in \cref{fig:patch_orig}). 
When providing only the original global context and generating local context and target patches from noise, we still observe good results, as we show in \cref{fig:patch_pipeline_whole}. 

\subsection{Experts Validation and Conditioned Image Generation}
\label{sec:experts}

To assess the realism of the images generated by MAMBO, we perform a user study with six radiologists, of whom three are juniors (currently pursuing their Ph.D.), and three are senior radiologists (who hold a Ph.D. with years of experience in clinical practice). They independently reviewed a total of 1,788 images (1,092 synthetic and 696 original), and were asked to identify the synthetic ones. The results are presented in \cref{tab:experts_validation}. 

\input{tables/experts_validation}

Performance varied significantly across evaluators, with no consistent pattern between seniority and recall/accuracy. Although the best-performing evaluator (Senior 0) achieved an F1 score of 0.867 and an accuracy of 0.849, Junior 2 also performed well (F1 score of 0.782 and accuracy of 0.752), outperforming two of the senior radiologists. With an overall recall value of 0.625 and an accuracy of 0.632 with a Wilson 95\% CI of [0.609, 0.654], the results indicate that it is challenging to distinguish MAMBO-generated images from real ones, even for experienced clinicians. This supports the visual fidelity and clinical plausibility of MAMBO's output. 

To evaluate MAMBO's controllability, the model was trained with conditioning on BI-RADS categories on RSNA data and assessed by expert radiologists. The study shows that MAMBO can generate realistic breast images representing both normal and pathological cases. Radiologists were able to identify key clinical findings, including masses and calcifications, and assign different BI-RADS categories. Upon expert evaluation, 22.9\% of synthetic images were labeled suspicious (BI-RADS 4 and 5), aligning with the original distribution presence of these anomalies, 28.6\%. This alignment suggests that MAMBO is not only capable of generating realistic lesions but also exhibits controllability over pathological content. 

\subsection{Classification}

A classification experiment is conducted to evaluate MAMBO-generated synthetic images for downstream tasks. The positive class contains images with a mass, while the negative class consists of images with no findings or BI-RADS categories 1 and 2. The train/test split follows the original partitioning from the VinDr dataset. From the training set, 20\% was randomly selected for validation.  Due to a highly unbalanced dataset (719 positive and 11,685 negative samples), an equal number of negative samples is randomly selected in each subset to ensure class balance. The final dataset sizes are 1,438 for training, 372 for validation, and 436 for testing.

\input{tables/classification}

The model from \cite{shen2019deep} is used as the classification backbone, as its input resolution of $1152\times896$ is similar to the mid-resolution image output of MAMBO. Separate models are trained with varying percentages of synthetic images, from 0\% to 100\%, replacing original samples. 

For this task, a Stage 1 model is trained using only the images containing a mass from the training split of the VinDr dataset. During model training, synthetic images are progressively introduced into the dataset. Specifically, for each successive model iteration, 10\% of the original images are removed from the positive and negative classes and replaced with synthetic images generated by the mid-resolution MAMBO model. All models are validated and tested using validation and test sets containing only real data. The results of the classification models are shown in \cref{tab:classification}.

 The model trained solely on original data achieved the highest AUC score of 0.744. However, it has a poor recall value of 0.404, which is one of the most important metrics for models used in medical studies, as one wants to miss as few positive instances as possible \cite{hicks2022evaluation}. In contrast, the model trained with a balanced mix of 50\% synthetic and 50\% original data achieves the highest recall of 0.752 and F1 score of 0.689, along with the second-best AUC score of 0.72. This proves that MAMBO's synthetic data, when combined with original data, can be used to enhance the training of these models. 

\subsection{Super-Resolution}

\input{images/sr_image_compare}

Image super-resolution (SR) refers to the task of reconstructing high-resolution (HR) images from their low-resolution (LR) counterparts \cite{wang2020deep}. In this work, SR is explored as a way to generate HR mammograms from the LR images, which can be synthesized using a standard low-resolution DDPM. Since no existing SR approaches were specifically developed for mammography, MAMBO was compared against bicubic interpolation and three general-purpose state-of-the-art SR models: AuraSR \cite{gigagan}, DDNM \cite{ddnm}, and LDM-SR \cite{rombach2022high}.

AuraSR is used as-is, without any additional training. DDNM, in contrast, was pretrained on mammography images. Two versions of the LDM-SR model were evaluated: the original model, pretrained on the OpenImages dataset, and a second version, pretrained on mammography images, to adapt the model to the target domain. In the latter case, the original LDM-SR training procedure is followed for training both the first-stage autoencoder model (with VQ regularization) and the second-stage diffusion model (U-Net). Separate models were trained for the RSNA and VinDr datasets.

All metrics were calculated using real mammograms. Most SR models cannot upscale to arbitrary resolutions; therefore, all images were first zero-padded to a uniform size of $4096\times4096$ pixels to preserve the resolution of the breast tissue. These padded images were then downscaled to $256 \times 256$ pixels to serve as inputs for the SR models.

Due to the high computational demands of DDNM, LDM-SR, and MAMBO, generating full-resolution outputs at scale was not feasible. As a workaround, models were evaluated using independent image patches. Each $256 \times 256$ input was first upscaled by a factor of 4 to generate a $1024 \times 1024$ intermediate image. From this, a $256 \times 256$ patch was extracted and then upscaled by a factor of 4. The final evaluation patch was obtained by center-cropping the resulting $1024 \times 1024$ image.

Quantitative results presented in \cref{tab:comparison_rsna} and \cref{tab:comparison_vindr} show performance on patches extracted from images upscaled by $4\times$ and $16\times$ for the RSNA and VinDr datasets, respectively. Metrics were computed by comparing 2,540 output patches from each model against their corresponding original patches at the same spatial location. Visual comparisons of output patches for both datasets are shown in \cref{fig:sr_comparison}.

\input{tables/mambo_compare}

MAMBO outperforms all other methods on the RSNA dataset at the $16\times$ scale, achieving the best FID (21.82) and LPIPS (0.31), and showing notably better performance than the second-best, LDM-SR (Mammo), which scores 144.65 and 0.45, respectively. On the same scale for the VinDr, MAMBO delivers the best LPIPS and the second-best FID. At $4\times$, AuraSR slightly outperforms MAMBO and LDM-SR (Mammo), but its performance degrades significantly at higher scales. Overall, MAMBO demonstrates the strongest performance, particularly at the native resolution used in digital mammography, delivering superior perceptual quality and fidelity.

\subsection{Data Memorization}
\label{sec:memorization}

One of the critical concerns in generating synthetic medical data is the risk of memorization and data leakage. Ensuring that synthetic data does not retain or reveal identifiable information is crucial for its safe and ethical use in medicine. 

To assess potential memorization, synthetic images produced by MAMBO were compared against real images from the training dataset to identify any significant visual overlap or unintended reproduction. Top-k nearest neighbors were retrieved using cosine similarity in the Mammo-CLIP embedding space \cite{ghosh2024mammo}. As shown in ~\cref{fig:memorization}, even when embedding similarity is high, the generated images differ clearly from the originals in shape, tissue structure, and texture. This indicates that MAMBO's outputs reflect genuine generalization, not memorization or data leakage. 

\input{images/figures/memorization}

\subsection{Anomaly Segmentation} 

The InBreast dataset was used to evaluate the anomaly segmentation performance. The evaluation scenario is basically zero-shot, as the images tested on are different from those used to train MAMBO on. To allow comparison with existing approaches used in the domain of radiology, only the low-resolution (global context) model was used as described in \cref{sec:method} and trained on the subset of healthy images taken from the VinDr dataset. 

To select the optimal value of the timestep parameter ($\lambda$) for the anomaly segmentation task, values in the $[300, 900]$ range were evaluated. The optimal value of $\lambda^*=700$, was obtained for all masses in the dataset, corresponding to an IoU of 0.216.

\begin{figure}[htp]
    \centering
    \begin{subfigure}[b]{0.11\textwidth}
        \includegraphics[width=\textwidth]{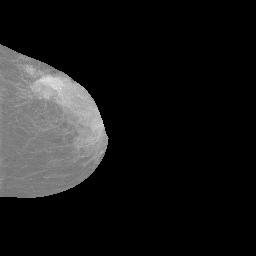}
        \caption{input}
        \label{fig:anomaly_original}
    \end{subfigure}\hspace{0.1cm}%
    \begin{subfigure}[b]{0.11\textwidth}
        \includegraphics[width=\textwidth]{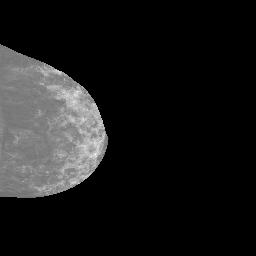}
        \caption{denoised}
        \label{fig:anomaly_denoised}
    \end{subfigure}\hspace{0.1cm}%
    \begin{subfigure}[b]{0.11\textwidth}
        \includegraphics[width=\textwidth]{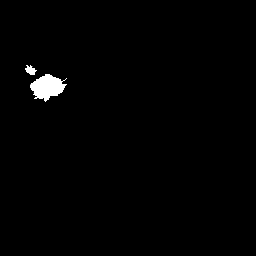}
        \caption{GT mask}
        \label{fig:anomaly_mask}
    \end{subfigure}\hspace{0.1cm}%
    \begin{subfigure}[b]{0.11\textwidth}
        \includegraphics[width=\textwidth]{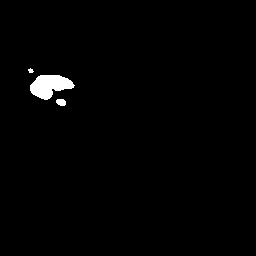}
        \caption{prediction}
        \label{fig:anomaly_prediction}
    \end{subfigure}\hspace{0.1cm}%

    \caption{Anomaly segmentation using the Stage 1 model.}
    \label{fig:anomaly_images}
\end{figure}

To further probe the capabilities of recognizing lesions of different sizes, \cref{tab:anomaly_size} illustrates the performance of MAMBO using the optimal $\lambda^*$. To produce the data, lesions are grouped in buckets according to size. The size of lesions (in pixels) in the dataset follows an exponential distribution. Lesions are assigned to the buckets using a logarithmic scale, which yields buckets with the same number of images (18) in each, except for the last bucket, which contains 17 images. 

\input{tables/anomaly_size}

As \cref{tab:anomaly_size} shows, larger lesions are easier to detect using the $256\times256$ model. Exceptions occur when a lesion grows large, nearing breast size, and the model finds it difficult to remove the whole anomaly. Quantitatively, the best result achieved by the proposed approach is as high as $0.423$, a significant improvement over the $0.269$ reported for AnoDDPM in ~\cite{wyatt2022anoddpm} for brain radiology images. 

\subsection{Ablation Studies}\label{sec:ablation}

Ablation studies are conducted to understand better how the number of channels in the U-Net model, as well as the size of the local context (controlled by the parameter $k$) affects the quality of the final image. As a baseline, a single unified model (Unified 4x) is trained to generate both local context and patches. Variations of the MAMBO model are explored as well: a model without global context, but with $k=3$ (MAMBO w/o gl. ctx.), as well models with the complete MAMBO architecture but with $k=1$ (Local ctx 256), $k=2$ (Local ctx 512) and $k=3$ (MAMBO).
\cref{tab:ablation} shows that \textit{Local ctx 256} performs the best quantitatively, followed by MAMBO. Images generated by \textit{Local ctx 256}, however, exhibit visible artifacts, making it not the best option. 

\input{tables/ablation}

%% file: tables/image_generation.tex
\begin{table}[!ht]
\caption{FID values obtained for MAMBO}
\label{tab:image_generation}
\centering
\small
{\resizebox{\columnwidth}{!}{
\begin{tabular}{>{\raggedright\arraybackslash}p{0.2\linewidth}>{\centering\arraybackslash}p{0.2\linewidth}>{\centering\arraybackslash}p{0.2\linewidth}>{\centering\arraybackslash}p{0.2\linewidth}>{\centering\arraybackslash}p{0.2\linewidth}}
\toprule
\textbf{Dataset} & \textbf{Input}        & \textbf{ROI} & \textbf{Random} & \textbf{Pipeline} \\
\midrule
\multirow{3}{*}{RSNA}  & patch       & 2.73  & 3.21  & 15.55 \\
                       & lc. context & 13.89 & 14.34 & 20.72 \\
                       & gl. context & N/A     & N/A     & 4.71  \\
\midrule
\multirow{3}{*}{VinDr} & patch       & 4.77  & 5.25  & 15.62 \\
                       & lc. context & 7.76  & 7.99  & 8.11  \\
                       & gl. context & N/A     & N/A     & 2.14  \\
\bottomrule
\multicolumn{5}{p{0.5\textwidth}}{All values are calculated on patches of size $256\times256$ pixels.}
\end{tabular}
}}
\end{table}

%% file: tables/experts_validation.tex
\begin{table}[htpb]
\centering
\caption{Medical experts validation}\label{tab:experts_validation}
{\resizebox{\columnwidth}{!}{

\begin{tabular}{lcccccc}
\toprule
& \textbf{Samples} & \textbf{Precision}  & \textbf{Recall}  & \textbf{F1} & \textbf{Accuracy} & \textbf{95\% CI (Wilson)} \\ \midrule
Junior 0 & 298 & 0.624 & 0.648 & 0.636 & 0.547 & [0.490, 0.602] \\
Junior 1 & 298 & 0.565 & 0.500 & 0.531 & 0.460 & [0.404, 0.517] \\
Junior 2 & 298 & 0.842 & 0.731 & 0.782 & 0.752 & [0.700, 0.797] \\
Senior 0 & 298 & 0.942 & 0.802 & 0.867 & 0.849 & [0.804, 0.885]\\
Senior 1 & 298 & 0.635 & 0.632 & 0.634 & 0.554 & [0.497, 0.609]\\
Senior 2 & 298 & 0.919 & 0.434 & 0.590 & 0.631 & [0.575, 0.684]\\ \midrule
Overall & 1788 & 0.733 & 0.625 & 0.675 & 0.632 & [0.609, 0.654] \\
\bottomrule
\multicolumn{7}{p{0.65\textwidth}}{Medical experts validation of synthetic images generated by MAMBO. Six radiologists evaluated 182 synthetic and 116 original images each to identify the synthetic ones.}
\end{tabular}
}}
\end{table}

%% file: tables/classification.tex
\definecolor{lightgreen}{rgb}{0.7, 0.98, 0.7}
\definecolor{lightblue}{rgb}{0.85, 0.85, 1}

\begin{table}[htpb]
\centering
\caption{Classifiers trained on MAMBO and original data}
\label{tab:classification}
{\resizebox{\columnwidth}{!}{
\begin{tabular}{ccccc}
\toprule

\textbf{Synthetic data (\%)} & \textbf{AUC} $\uparrow$ & \textbf{Recall} $\uparrow$ & \textbf{Precision} $\uparrow$ & \textbf{F1} $\uparrow$ \\ \midrule

0 &\textbf{0.744} & 0.404 & \textbf{0.863} & 0.550 \\ 
10  & 0.714 & 0.495 & 0.730 & 0.590 \\ 
20  & 0.718 & 0.583 & 0.747 & 0.655 \\
30  & 0.698 & \underline{0.633} & 0.690 & \underline{0.660} \\ 
40  & 0.710 & 0.550 & 0.710 & 0.620 \\
\rowcolor{lightblue}
50  & \underline{0.720} & \textbf{0.752} & 0.636 & \textbf{0.689} \\ 
60  & 0.715 & 0.477 & \underline{0.759} & 0.586 \\ 
70  & 0.702 & 0.495 & 0.720 & 0.587 \\ 
80  & 0.601 & 0.239 & 0.754 & 0.362 \\ 
90  & 0.619 & 0.317 & 0.697 & 0.435 \\ 
100   & 0.598 & 0.330 & 0.631 & 0.434 \\ 
\bottomrule
\multicolumn{5}{p{251pt}}{Evaluation of classification model trained with different proportions of synthetic images in the training dataset. The best values are highlighted in \textbf{bold}, and the second-best values are \underline{underlined} to emphasize performance differences.}
\end{tabular}}}

\end{table}

%% file: images/sr_image_compare.tex
\captionsetup[subfigure]{font=scriptsize} 

\begin{figure*}[!htpb]
    \centering
        \includegraphics[width=0.12\textwidth]{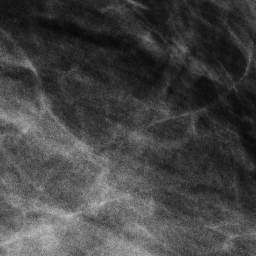}
        \includegraphics[width=0.12\textwidth]{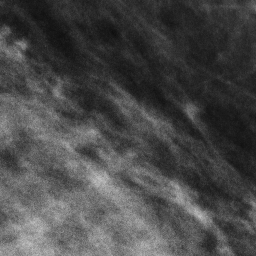}     
        \includegraphics[width=0.12\textwidth]{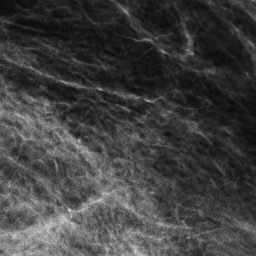}
        \includegraphics[width=0.12\textwidth]{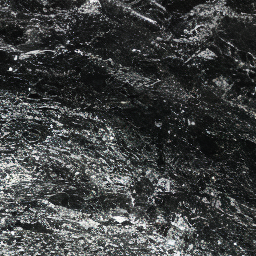}
        \includegraphics[width=0.12\textwidth]{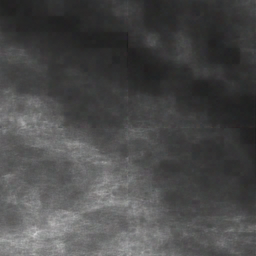}
        \includegraphics[width=0.12\textwidth]{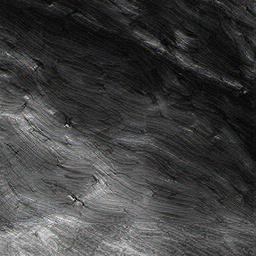}
        \includegraphics[width=0.12\textwidth]{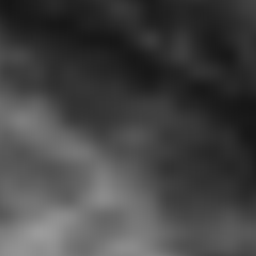}
        
\vspace{0.1cm}
\centering
    \begin{subfigure}[t]{0.12\textwidth}
        \includegraphics[width=\textwidth]{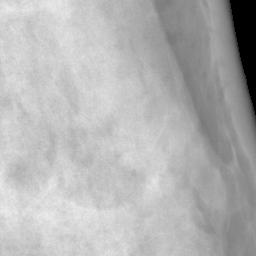}
        \caption{Original}
        \label{fig:vindr_patch_orig}
    \end{subfigure}
    \begin{subfigure}[t]{0.12\textwidth}
        \includegraphics[width=\textwidth]{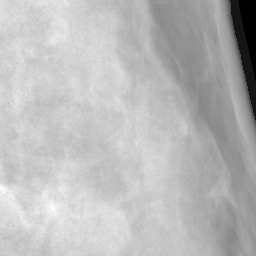}
        \caption{MAMBO\\}
        \label{fig:vindr_patch_mambo}
    \end{subfigure}
    \begin{subfigure}[t]{0.12\textwidth}
        \includegraphics[width=\textwidth]{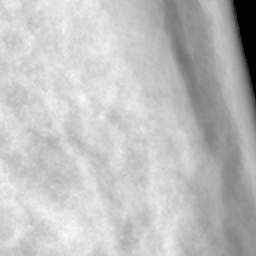}
        \caption{LDM-SR Mammo}
        \label{fig:vindr_patch_ldm_mammo}
    \end{subfigure}
    \begin{subfigure}[t]{0.12\textwidth}
        \includegraphics[width=\textwidth]{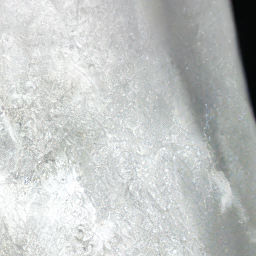}
        \caption{LDM-SR}
        \label{fig:vindr_patch_ldm_open_images}
    \end{subfigure}
    \begin{subfigure}[t]{0.12\textwidth}
        \includegraphics[width=\textwidth]{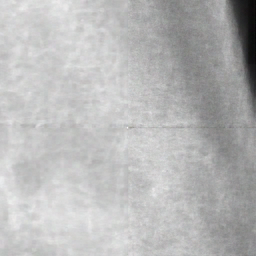}
        \caption{DDNM}
        \label{fig:vindr_patch_ddnm}
    \end{subfigure}
    \begin{subfigure}[t]{0.12\textwidth}
        \includegraphics[width=\textwidth]{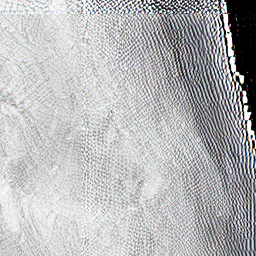}
        \caption{AuraSR}
        \label{fig:vindr_patch_aura}
    \end{subfigure}
    \begin{subfigure}[t]{0.12\textwidth}
        \includegraphics[width=\textwidth]{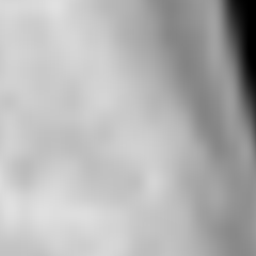}
        \caption{Bicubic}
        \label{fig:vindr_patch_bicubic}
    \end{subfigure}
    \caption{Qualitative results of different methods from RSNA (first row) and VinDr (second row) datasets for $16\times$ SR.
 }
    \label{fig:sr_comparison}
\end{figure*}

\captionsetup[subfigure]{font=footnotesize}

%% file: tables/mambo_compare.tex
\begin{table}[!ht]
\centering
\caption{Super-Resolution Comparison on RSNA}
\label{tab:comparison_rsna}
{\resizebox{\columnwidth}{!}{
\begin{tabular}{lcccc}
\hline
& \multicolumn{2}{c}{\textbf{4x}}  & \multicolumn{2}{c}{\textbf{16x}} \\ 
\cmidrule(lr){2-3}\cmidrule(lr){4-5} & \textbf{FID} $\downarrow$ & \textbf{LPIPS} $\downarrow$ & \textbf{FID} $\downarrow$ & \textbf{LPIPS} $\downarrow$ \\ 
\midrule
Bicubic                                   & 126.91 & 0.36 &  215.81  & 0.66  \\
AuraSR \cite{gigagan}                     &  \textbf{28.40} & \textbf{0.11} & 249.00  & 0.69  \\
DDNM (Mammo)\cite{ddnm}                   & 110.92 & \underline{0.20} & 227.21  &  0.53  \\
LDM-SR (OpenImages)\cite{rombach2022high} &  80.69 & 0.49 & 270.38  &  0.76 \\
LDM-SR (Mammo) \cite{rombach2022high}     & \underline{31.87}  & 0.25 & \underline{144.65}  &  \underline{0.45}  \\ 
MAMBO                                     & 33.70 & 0.25 & \textbf{21.82}   &  \textbf{0.31}  \\ 
\bottomrule
\multicolumn{5}{p{0.5\textwidth}}{LPIPS \cite{zhang2018unreasonable} used with the squeezeNet. The best values are highlighted in \textbf{bold}, and the second-best values are \underline{underlined}.}\\
\end{tabular}
}}
\end{table}

\begin{table}[!ht]
\centering
\caption{Super-Resolution Comparison on VinDr}
\label{tab:comparison_vindr}
{\resizebox{\columnwidth}{!}{
\begin{tabular}{lcccc}
\hline
& \multicolumn{2}{c}{\textbf{4x}}  & \multicolumn{2}{c}{\textbf{16x}} \\ 
\cmidrule(lr){2-3}\cmidrule(lr){4-5} & \textbf{FID} $\downarrow$ & \textbf{LPIPS} $\downarrow$ & \textbf{FID} $\downarrow$ & \textbf{LPIPS} $\downarrow$ \\ 
\midrule
Bicubic                                   & 74.96 & 0.21 &  240.71  & 0.41  \\
AuraSR \cite{gigagan}                     &  \textbf{17.48} & \textbf{0.07} & 257.28  & 0.71  \\
DDNM (Mammo)\cite{ddnm}                   & 99.03 & 0.13 & 219.62  &  0.40  \\
LDM-SR (OpenImages)\cite{rombach2022high} & 104.90 & 0.16 & 197.53  &  0.49  \\
LDM-SR (Mammo) \cite{rombach2022high}     &  18.94 & \underline{0.08}  & \textbf{19.17}   &  \underline{0.16}  \\ 
MAMBO                                     & \underline{18.74} & 0.10 & \underline{26.06}   & \textbf{0.13}  \\ 
\bottomrule
\multicolumn{5}{p{0.5\textwidth}}{LPIPS \cite{zhang2018unreasonable} used with the squeezeNet. The best values are highlighted in \textbf{bold}, and the second-best values are \underline{underlined}.}\\
\end{tabular}
}}
\end{table}

%% file: images/figures/memorization.tex
\begin{figure}[htb!]
    \centering
        \includegraphics[width=0.9\columnwidth]{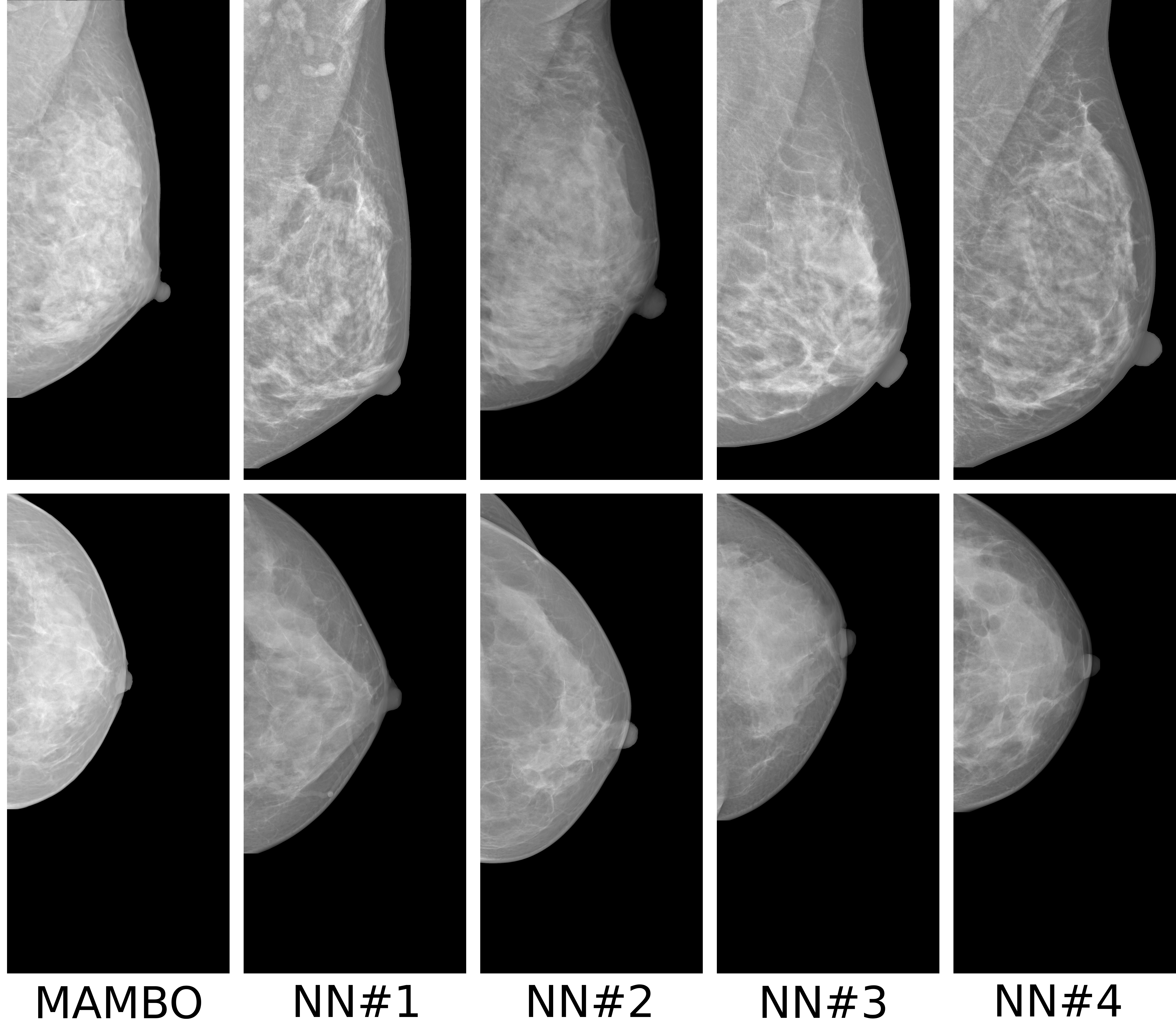}
    \caption{Images generated by MAMBO alongside their top-4 nearest neighbors from the training set.
 }
    \label{fig:memorization}
\end{figure}

%% file: tables/anomaly_size.tex
\begin{table}[!t]
\centering
\caption{Average IoU by lesion size groups for $\lambda^*$ = 700}
\label{tab:anomaly_size}
{\resizebox{\columnwidth}{!}{
\begin{tabular}{lcccccc}
\toprule
\textbf{Area (px)} & 64 & 128 & 222 & 521 & 951 & 2187 \\
\midrule
\textbf{IoU} $\uparrow$ & 0.019 & 0.069 & 0.125 & 0.280 & \textbf{0.423} & 0.393 \\
\bottomrule
\multicolumn{7}{p{0.99\columnwidth}}{Area is represented on a log scale, median size is shown for each bucket.}
\end{tabular}
}}
\end{table}

%% file: tables/ablation.tex
\begin{table}[htp]
    \centering
    \caption{Ablation studies}    
    \label{tab:ablation}
    {\resizebox{\columnwidth}{!}{
    \begin{tabular}{lcc cc cc}
        \toprule
         &  &  & \multicolumn{2}{c}{\textbf{Local}} & \multicolumn{2}{c}{\textbf{Patch}} \\  
        \cmidrule(lr){4-5}\cmidrule(lr){6-7}
        \textbf {Model}  & \textbf {\# channels} & \textbf {k} & \textbf{FID} $\downarrow$ & \textbf{LPIPS} $\downarrow$ & \textbf{FID} $\downarrow$ & \textbf{LPIPS} $\downarrow$ \\
        \midrule
        Unified 4x & 2   &  4   & 34.69 & 0.49 & 51.84 & 0.39  \\
        MAMBO w/o gl. ctx & 2   &  3  & 19.78 & 0.48 & 35.31  &  0.35 \\
        Local ctx 256    & 3 &   1   & N/A & N/A & 29.96   &  0.34 \\
        Local ctx 512    &  3 & 2    & 35.92 & 0.50 & 36.46 &    0.38   \\
        Local ctx 768 (MAMBO) & 3 &  3  & 21.57 & 0.48 & 34.55  &  0.34 \\
        \bottomrule    
        \multicolumn{7}{p{251pt}}{Metrics on local contexts and patches with different models}
    \end{tabular}}}

\end{table}

%% file: sec/5_conclusion.tex
\section{Conclusions and Future work}
\label{sec:conclusion}
This paper presents MAMBO, a novel patch-based denoising diffusion generative approach capable of producing full-resolution digital mammograms. By leveraging separate diffusion models for global and local contexts, semantically plausible image generation is achieved while minimizing boundary effects. This innovative strategy enables generation of mammograms with a resolution of $3840\times3840$ pixels, marking significant improvements over existing methods. Additionally, MAMBO can be utilized for anomaly segmentation, allowing pixel-wise predictions on lesion presence without the need to be trained on pixel-wise labels, as well as to enhance the training of classification models designed with mammography applications in mind. 
Experiments demonstrated state-of-the-art performance in image generation, super-resolution and anomaly segmentation, underscoring the potential to improve patient outcomes through improved mammography analysis accuracy and earlier lesion detection. Beyond numerical verification, expert radiologists conducted thorough evaluations, affirming its clinical relevance and practical applicability in real-world medical imaging scenarios.
Looking forward, user guidance will be integrated into the diffusion process, by utilizing metadata or text features from mammogram descriptions. Furthermore, efforts will focus on detecting small and emerging anomalies using high-resolution pipelines.

%% file: sec/X_suppl.tex
\clearpage

\setcounter{page}{1}

\maketitlesupplementary

We begin by providing additional details about the datasets used in our quantitative evaluation in \cref{supp:datasets}, then we describe the metrics used in \cref{supp:metrics}. Implementation details are provided in \cref{supp:impl_details}, and we describe the preprocessing steps in \cref{sec:tissue_mask}. We showcase the role of the overlap size in \cref{sec:overlap_patches}. \cref{supp:super_res} provides details about the super-resolution baselines, \cref{supp:anomaly} shows additional results on the anomaly segmentation task at varying $\lambda$ noising steps, and \cref{supp:ablation} describes more in details the implementation details of the ablation study. An ethical statement is provided in \cref{supp:ethic}. The remaining sections show additional qualitative results, in particular examples of network inputs in \cref{supp:net_input}, additional generated samples in \cref{supp:additional_samples}, and expert annotations on synthetic data in \cref{supp:expert_annotation}.

\section{Datasets}
\label{supp:datasets}

\noindent\textbf{VinDr} The VinDr dataset consists of 20,000 full-field digital mammogram images in DICOM format, 2,254 of which are annotated to contain different kinds of lesions, while the rest are labeled as healthy. Since the number of images with lesions is insufficient to train the model to reproduce such structures, the training is restricted to healthy images, which also allows to conduct the anomaly segmentation experiments. Additionally, only images taken by Siemens Mammomat Inspiration are considered, which are of $3518\times2800$-pixel resolution and comprise the majority of healthy images (13,942 out of 17,746). This dataset contains images in CC and MLO views, diverse in terms of breast density.

\noindent\textbf{RSNA} This dataset has been collected by the Radiological Society of North America (RSNA) and contains 54,706 images captured by 10 different machines. Data captured by a single machine, the one with the largest number of samples (ID 49) has been used in the study. The resolution of these images is $4096\times3328$, with 14,898 images in the CC and MLO views, diverse in terms of the type of tissue.

\noindent\textbf{InBreast} This dataset contains 410 full-field digital mammograms with a resolution of $3328\times2560$ pixels. Pixel-wise labels are provided on mammograms where lesions are present, allowing to benchmark MAMBO in terms of the localization of anomalies. There are different types of lesions present in the mammograms, with the focus here being on masses, which are present in 107 images. Other types of lesions, \eg asymmetries, distortions and calcifications, are not of interest for analysis as they occupy a significant portion of the whole breast and are easily detectable even for traditional methods. Masses, on the other hand,  are compact in space, with variable size and need to be precisely distinguished from healthy tissue. 

\section{Metrics} 
\label{supp:metrics}

\begin{itemize}
    \item \textbf{FID} is widely used to assess the quality of images created by a generative model, by comparing the distributions of downstream neural network activations of the compared datasets. The higher the quality of the synthetic dataset, the lower the expected FID score when compared to real images.
    \item \textbf{LPIPS} is another DL-based measure, capturing perceptual similarity between images, the higher value indicating that the images differ more.
    \item \textbf{IoU} represents the ratio between the intersection of two shapes and their union, taking the value of 1 for the ideal overlap and 0 if there is no intersection. 
    \item \textbf{AUC} measures the model's ability to distinguish between diseased and non-diseased samples. 
    \item \textbf{Recall} evaluates the ability of the model to identify diseased instances correctly. 
    \item \textbf{Precision} evaluates the ability of the model to identify non-diseased instances correctly.
    \item \textbf{F1} aggregates recall and precision in an integral metric.
\end{itemize}

\section{MAMBO Implementation Details} 
\label{supp:impl_details}

As a preprocessing step, all original images are normalized so that the pixel values fit the $[0,1]$ range. The images are pre-processed to neutralize the negative where needed, removing all artifacts and keeping only the breast in the image (\ie, making the background totally black) and padding the image to a square on the opposite side of the breast. Images where the breast is on the right side are flipped to make the whole dataset consistently contain only the breasts on the left side. After this preprocessing, the resulting resolution of images from the RSNA dataset is $3769\times3769$ pixels, while VinDr images have a resolution of 3237 pixels on one side. Additionally, images are padded to the closest width and height divisible by 256 to keep them from getting deformed when scaled down. A binary segmentation mask of the whole breast is created to crop patches containing breast tissue. The random coordinate is selected from the pixels contained on the segmentation mask and used as the center of the patch. In this way, it is ensured that the training patches always contain important data that is useful for training.

All diffusion models are based on the PyTorch implementation of DDPM \cite{diffusion}\footnote{https://github.com/lucidrains/denoising-diffusion-pytorch} with a change of the noising/denoising procedures.  For all models, the noise sampling is conducted using a linear noise scheduler for $1000$ diffusion timesteps. Training is done with the Adam optimizer, a learning rate of $5\times 10^{-5}$, and a batch size of $8$. The training is performed on NVIDIA A100 GPUs. The global-context model used to generate whole mammograms is trained for 124,500 iterations on 2 GPUs. The training of the model for patch generation takes 170,000 iterations on 4 GPUs. Finally, this model is fine-tuned for 31,000 iterations on 4 GPUs to get the model for the generation of local contexts. The parameter details are specified in \cref{tab:hyperparameters}

\input{tables/model_hyperparameters}

The three-channel models have $78.25M$ parameters, while the single-channel global-context model has slightly less parameters ($78.24M$), which is at the lower end of the state-of-the-art patch-based diffusion models' size spectrum \cite{neighbour-patch}. Patch-DM in its full form consists of $154M$ parameters, while the full MAMBO ensemble has $\sim235M$ parameters, which is smaller than all models Patch-DM compares with.

\section{Image Preprocessing}
\label{sec:tissue_mask}
\begin{figure*}
    \centering
    \begin{subfigure}[b]{0.18\textwidth}
        \includegraphics[width=\textwidth]{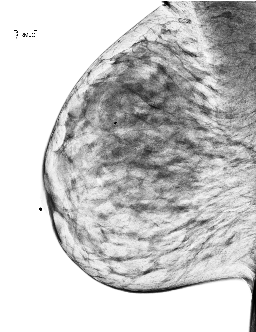}
        \caption{original}
    \end{subfigure}\hspace{0.1cm}%
    \begin{subfigure}[b]{0.18\textwidth}
        \includegraphics[width=\textwidth]{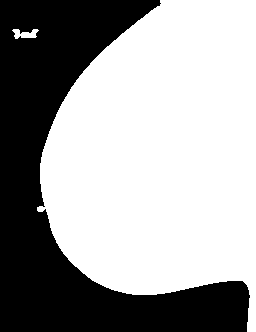}
        \caption{initial mask}
    \end{subfigure}\hspace{0.1cm}%
    \begin{subfigure}[b]{0.18\textwidth}
        \includegraphics[width=\textwidth]
        {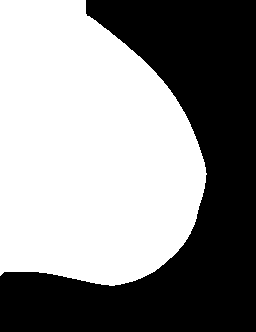}
        \caption{largest blob}
    \end{subfigure}\hspace{0.1cm}%
    \begin{subfigure}[b]{0.18\textwidth}
        \includegraphics[width=\textwidth]
        {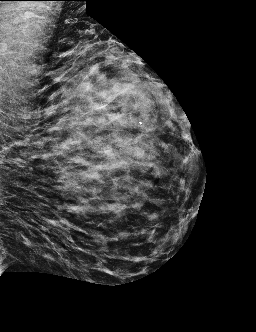}
        \caption{final result}
    \end{subfigure}\hspace{0.1cm}%

    \caption{
    The preprocessing applied to images before they could be used for the anomaly segmentation.}
    \label{fig:preprocessing}
\end{figure*}

Several steps of preporcessing were performed on the images, before they were used for training or as input to the anomaly segmentation pipeline. If an original image was ''inverted'', i.e. background was white and the breast tissue indicated in shades of grey, the image was first inverted to create standard white-breast-tissue on black background mammograms. Automatic segmentation of the breast was then performed and the values of the pixels outside the segmentation set to zero. If the segmentation mask indicated that the breast is located on the right side of the frame, the image was flipped horizontally. The process is illustrated in Fig. \ref{fig:preprocessing}. 

The mask, corresponding to the inner tissue of the breast, was created by applying a Gaussian blur filter, binarizing the original image, taking the largest blob and eroding it.  

The dimensions of the Gaussian kernel were set to $\sim 50\times$ less than the image size. Thresholding of the blurred image was then performed to create a binary image. This threshold was set by multiplying the Otsu threshold of the image by $0.175$. This lowered variant of the Otsu threshold, determined experimentally, managed to include into the breast mask all the dark (non-dense) parts of the tissue, which would otherwise be missed. The background shades, represented by the lower percentiles of the pixel-intensity distribution, were dark enough not to get spuriously included. 

On such, binarized images, contours were found and the largest one in terms of area selected for further processing. This blob is then eroded using a $20\times20$ kernel the boundary of the breast. This helps isolate true anomalies, as the diffusion process has been shown to sometimes introduce unwanted changes in the breast's boundary. This is understandable, when one considere that this part is not as homogeneous as the rest of the tissue. 

As no breast masks are provided in the dataset, all the parameters of the different preprocessing steps were determined manually and subsequently evaluated. 

\section{Overlapping Patches for Smooth Transition}
\label{sec:overlap_patches}

\begin{figure}[htpb!]
\centering
\includegraphics[width=0.45\textwidth]{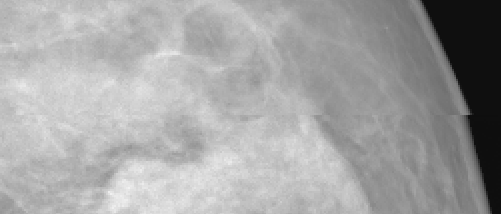}
\includegraphics[width=0.45\textwidth]{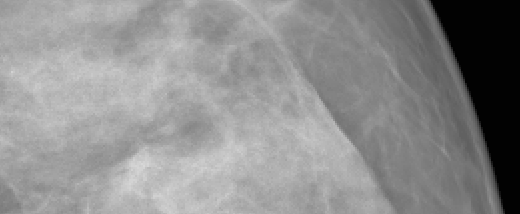}
\caption{
Crop of the region where two local contexts are stitched together, before (first row) and after (second row) applying the overlap strategy described in Subsec. \ref{section:inference} to smooth the transitions when stitching. The first approach generated every patch independently, without overlapping pixels, and then combined them to create the final image. This resulted in visible artifacts (first row) in the places where patches were stitched. Once the 32-pixel overlap with the neighboring, previously-generated patches is applied, the checkerboard effect and additional discontinuity artifacts disappear (second row).
}
\label{fig:overlap}
\end{figure}

Fig. \ref{fig:overlap} illustrates the effect of the autoregressive approach to generating patches, which was employed to address the chekerboard artifacts. When generating a patch, overlap (32-pixel wide), with the previously generated patches was used to enforce continuity, both on the left and top side of the patch being generated.

\begin{figure}[htbp]
    \centering
    \includegraphics[width=0.45\textwidth]{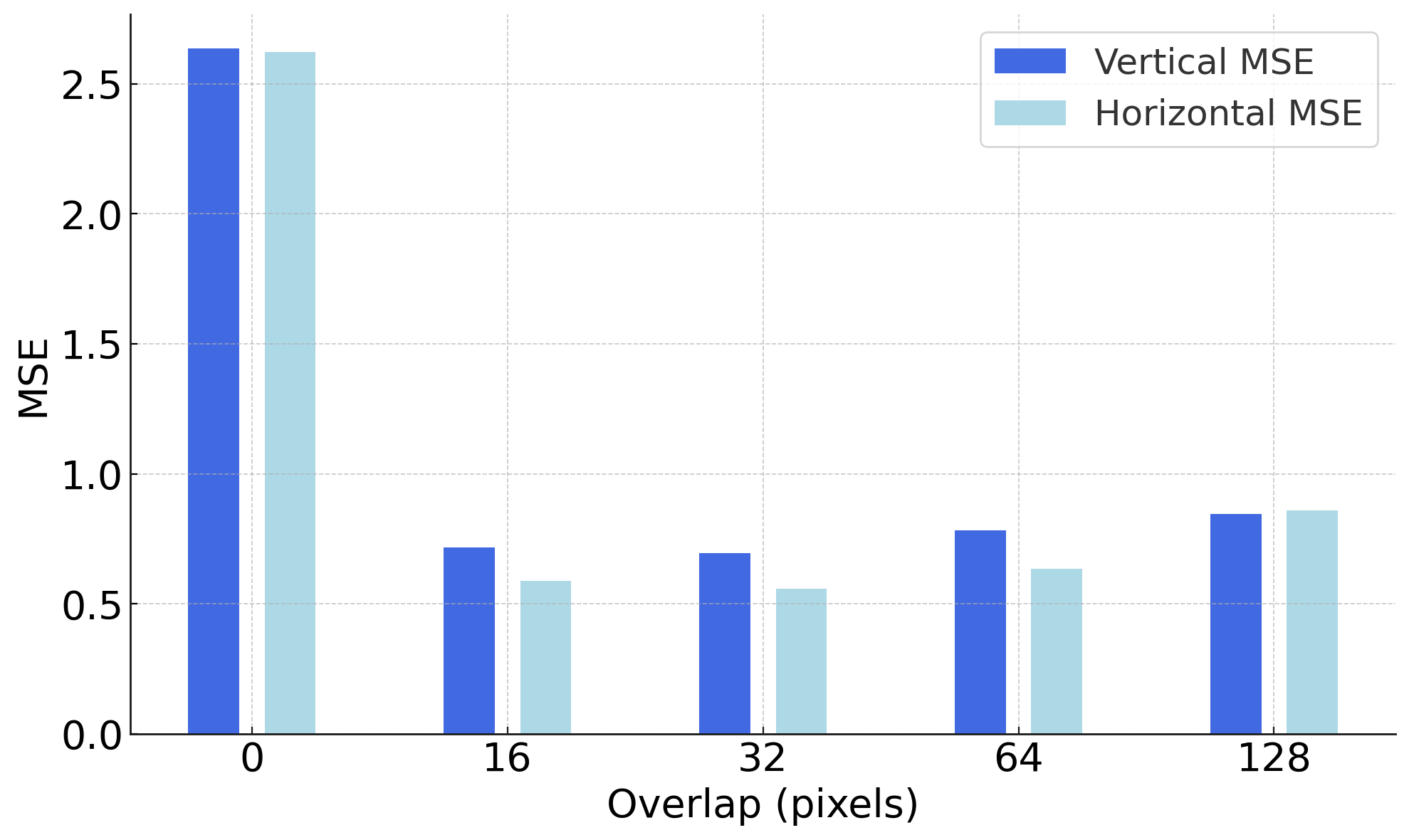}
    \caption{MSE on neighboring pixels on stitching boundaries  at varying size of the overlap.}
    \label{fig:mse_sticthing}
\end{figure}

To determine the optimal amount of overlap, 140 images are generated using the MAMBO pipeline with different values for the overlap as shown in \cref{fig:mse_sticthing}. 
The Mean-square error (MSE) is calculated between the neighboring pixels on the stitching boundaries, vertical and horizontal, to check whether there is a significant change in pixel values, which would indicate the presence of an artifact. It turns out that any of the overlaps used can drastically improve the blending, which can be confirmed both visually and by the drop in MSE. 32-pixel overlap is chosen for the experiments as it shows a slightly better MSE score than others and keeps the computation requirements low.

\section{Super-Resolution Baselines}
\label{supp:super_res}

AuraSR is a GAN-based super-resolution model designed for real-world applications. It is based on the GigaGAN \cite{gigagan} architecture and focuses on generating high-quality super-resolved images by adapting convolutional layers and attention mechanisms to handle complex image distributions. 

DDNM (Denoising Diffusion  Null-Space Model) \cite{ddnm} is a diffusion-based model that generates high-quality images through a gradual denoising process. By reversing a learned noise distribution, it transforms random noise into coherent images, making it a strong candidate for image generation tasks. DDNM was pretrained with mammography images to enable a fair comparison.



\section{Additional Results for Anomaly Segmentation on the InBreast Dataset}
\label{supp:anomaly}

Tab. \ref{tab:buckets} shows the IoU score for the anomaly segmentation sub-task as a function of the number of timesteps $\lambda$ used to add noise to the image. For $\lambda$ values less than 600, model performance in terms of the IoU score is low. This could mean that the level of noise, for those values of $\lambda$, is still too low to cover the target anomaly completely.

\input{tables/lambda}

\begin{figure*}[htp]
    \centering
    
    \includegraphics[width=0.2\textwidth]{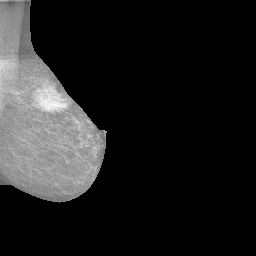}\hspace{0.1cm}%
    \includegraphics[width=0.2\textwidth]{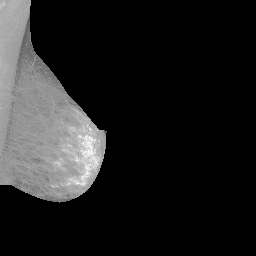}\hspace{0.1cm}%
    \includegraphics[width=0.2\textwidth]{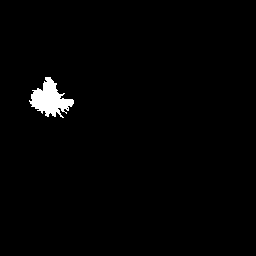}\hspace{0.1cm}%
    \includegraphics[width=0.2\textwidth]{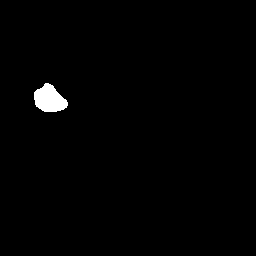}\hspace{0.1cm}%

    \includegraphics[width=0.2\textwidth]{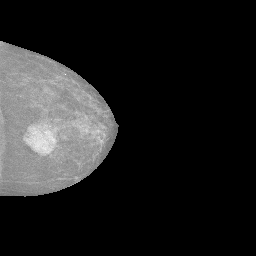}\hspace{0.1cm}%
    \includegraphics[width=0.2\textwidth]{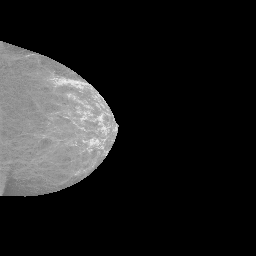}\hspace{0.1cm}%
    \includegraphics[width=0.2\textwidth]{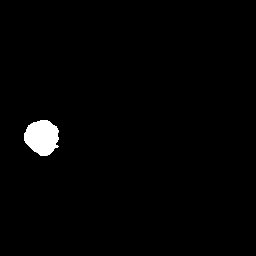}\hspace{0.1cm}%
    \includegraphics[width=0.2\textwidth]{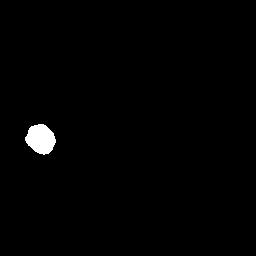}\hspace{0.1cm}%
    
    \includegraphics[width=0.2\textwidth]{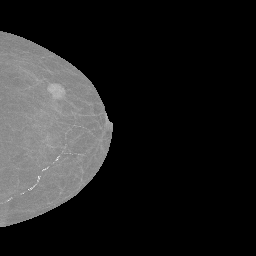}\hspace{0.1cm}%
    \includegraphics[width=0.2\textwidth]{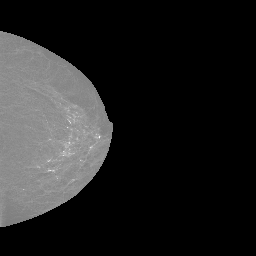}\hspace{0.1cm}%
    \includegraphics[width=0.2\textwidth]{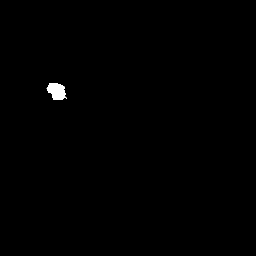}\hspace{0.1cm}%
    \includegraphics[width=0.2\textwidth]{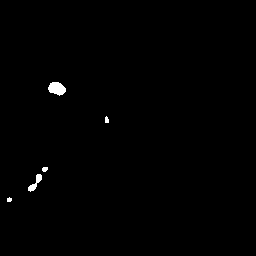}\hspace{0.1cm}%

    \begin{subfigure}[b]{0.2\textwidth}
        \includegraphics[width=\textwidth]{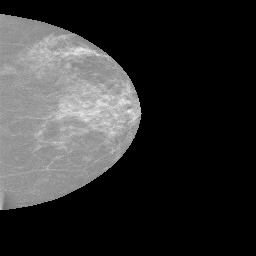}
        \caption{input image}
    \end{subfigure}\hspace{0.1cm}%
    \begin{subfigure}[b]{0.2\textwidth}
        \includegraphics[width=\textwidth]{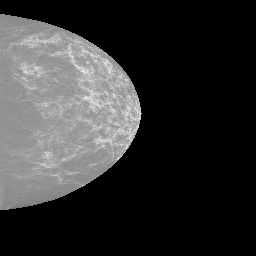}
        \caption{denoised image}
    \end{subfigure}\hspace{0.1cm}%
    \begin{subfigure}[b]{0.2\textwidth}
        \includegraphics[width=\textwidth]{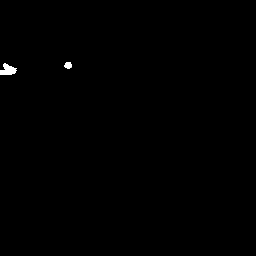}
        \caption{ground truth}
    \end{subfigure}\hspace{0.1cm}%
    \begin{subfigure}[b]{0.2\textwidth}
        \includegraphics[width=\textwidth]{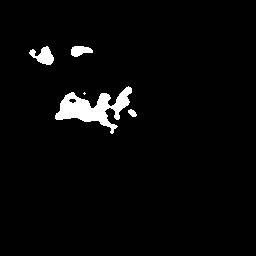}
        \caption{prediction mask}
    \end{subfigure}\hspace{0.1cm}%

    \caption{Sample results of anomaly segmentation for $\lambda=700$. The lesion mask in the first row, with an area of 938 pixels, falls into the optimal bucket (Tab. \ref{tab:buckets}) centered around 951 pixels, making it easier for the model to detect it. In contrast, the mask in the last row, with an area of 176 pixels, falls into the bucket centered around 128 pixels, where the model struggles to separate it from the spurious diffusion-induced changes.
    }
    \label{fig:suppl_anomaly_images}
\end{figure*}

Figure \ref{fig:suppl_anomaly_images} shows additional results of anomaly segmentation on the InBreast dataset, using the MAMBO low-resolution (global-context) model, for the size of lesion that the approach detects well, as well as one which is challenging for it. 

\section{Ablation Studies}
\label{supp:ablation}

Each model generated 400 mid-resolution and full-resolution images, from which a random 5100 local contexts and patches, $256\times256$ pixels, are cropped for further analysis. These images are used to compute FID and LPIPS in comparison to the original patches from the dataset. DDIM \cite{song2020denoising}, with 150 steps, is used for all models to accelerate the image generation process. 

The model in the first row in \cref{tab:ablation}, labeled \textit{Unified 4x}, is trained to upscale the input images of an arbitrary level of details by $4\times$ in both dimensions, so that the final images are generated in two stages using a single U-Net model. The mid-resolution image obtained after the first stage is resized from $1024\times1024$ to the resolution of $768\times768$ to be able to get the desired resolution of $3072\times 3072$ in the final step. This model shows the poorest performance in terms of FID and LPIPS metrics. However, the advantage is that it can be used for generating images of arbitrary resolutions without providing additional guidance.

The model labeled \textit{MAMBO w/o gl. ctx.} in \cref{tab:ablation} utilizes the Stage 3 MAMBO model, while the second stage model is a modified Stage 2 MAMBO  model, with  the first channel, \ie global context, removed. Quantitatively, this model shows metric values similar to MAMBO, but the generated images have visible artifacts shown in \cref{fig:ablation_bad}.

The model with the best FID and LPIPS scores of 29.96 and 0.34, respectively, labeled as \textit{Local context 256},  generates full-resolution images from initial low-resolution ones in one step. The model is trained using a 3-channel U-Net model similar to the Stage 2 MAMBO model, but different in terms of the resolution of the output patch. Although this model shows the best quantitative performance, it struggles to generate images without artifacts; therefore, this model was not selected as the best option. Examples of the artifacts in the images generated with this model are shown in \cref{fig:ablation_bad}.

The model labeled \textit{Local context 512} in \cref{tab:ablation} is the one most similar to the MAMBO. The only difference is that the parameter $k$ is set to 2, \ie local context represents a crop of $512\times 512$ pixels. \cref{tab:ablation} shows that MAMBO only slightly outperforms this model, indicating that the model is not significantly influenced by the resolution of the local context. 

\section{Ethical Statement}
\label{supp:ethic}

All the datasets used in the article are fully anonymized, open to the public for research purposes, approved by relevant oversight boards, and thoroughly exploited in the literature. 

As the method utilizes only anonymized image data and does not involve personal patient information, no risks have been detected in the context of ensuring data privacy. 

The research in the article makes a step towards more effective cancer diagnostics, creating a potential for numerous societal benefits.

\clearpage
\onecolumn

\section{Per-channel Input Examples}
\label{supp:net_input}

\begin{figure*}[htpb!]
    \centering
    \begin{subfigure}[b]{0.22\textwidth}
        \includegraphics[width=\textwidth]{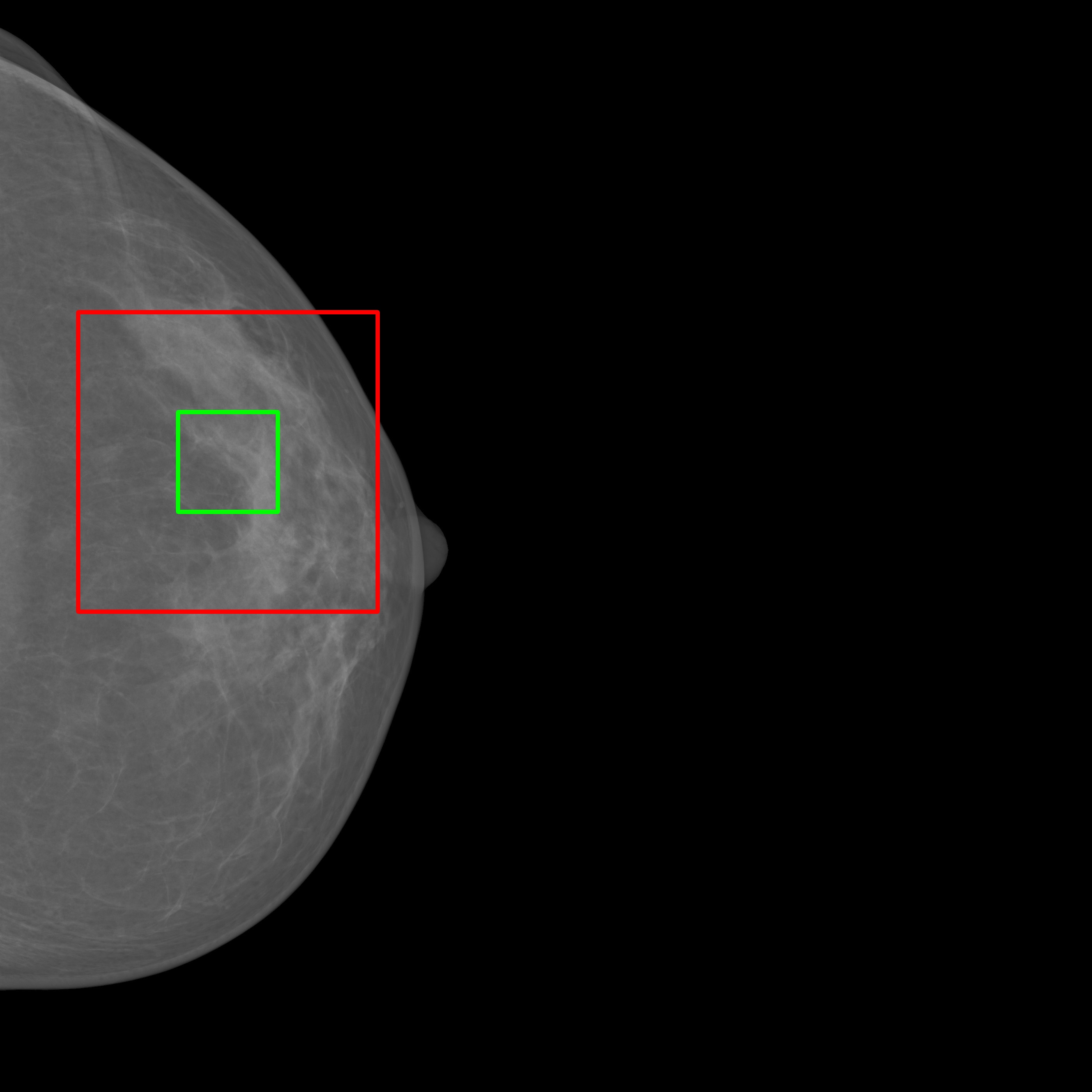}
    \end{subfigure}\hspace{0.1cm}%
    \begin{subfigure}[b]{0.22\textwidth}
        \includegraphics[width=\textwidth]{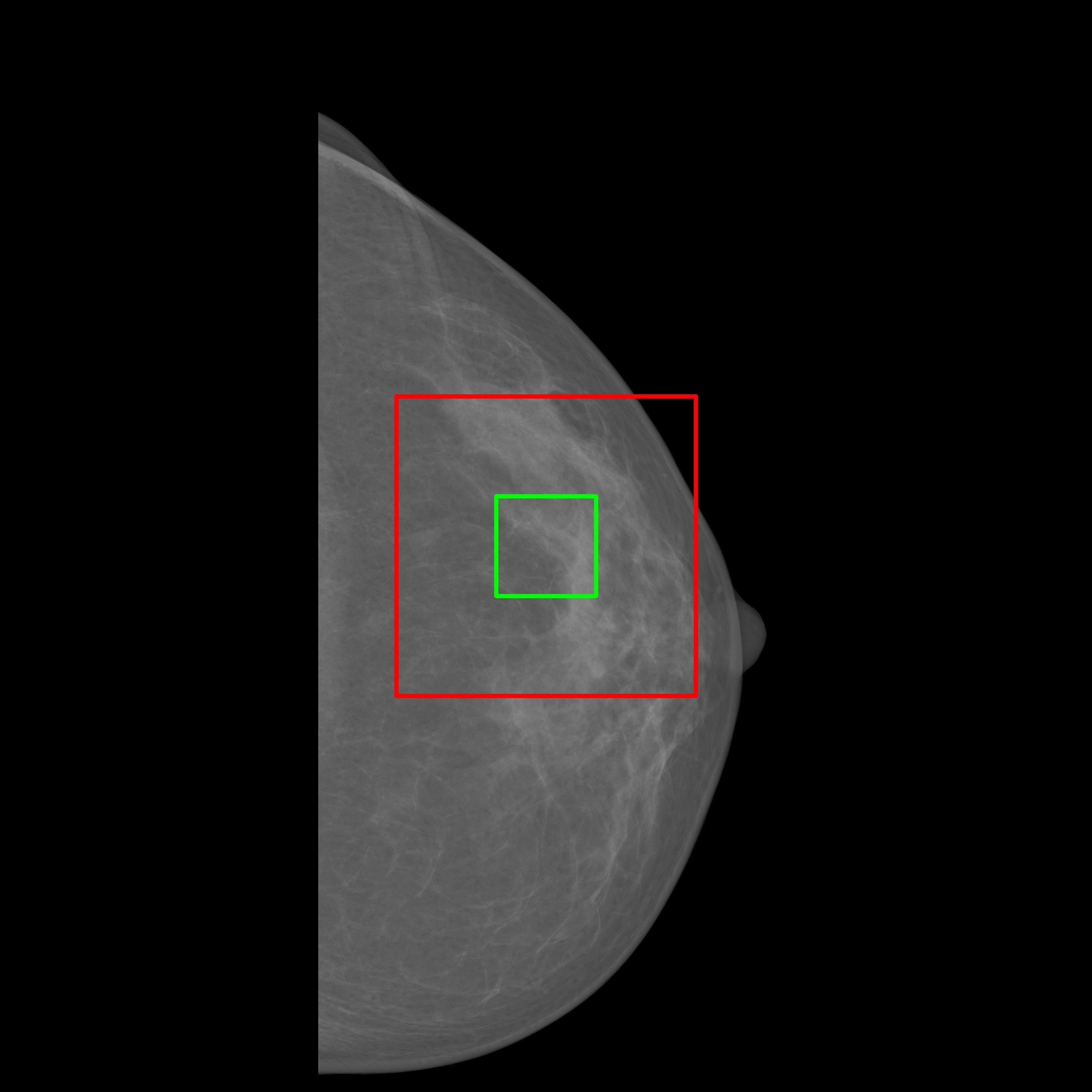}
    \end{subfigure}\hspace{0.1cm}%
    \begin{subfigure}[b]{0.22\textwidth}
        \includegraphics[width=\textwidth]{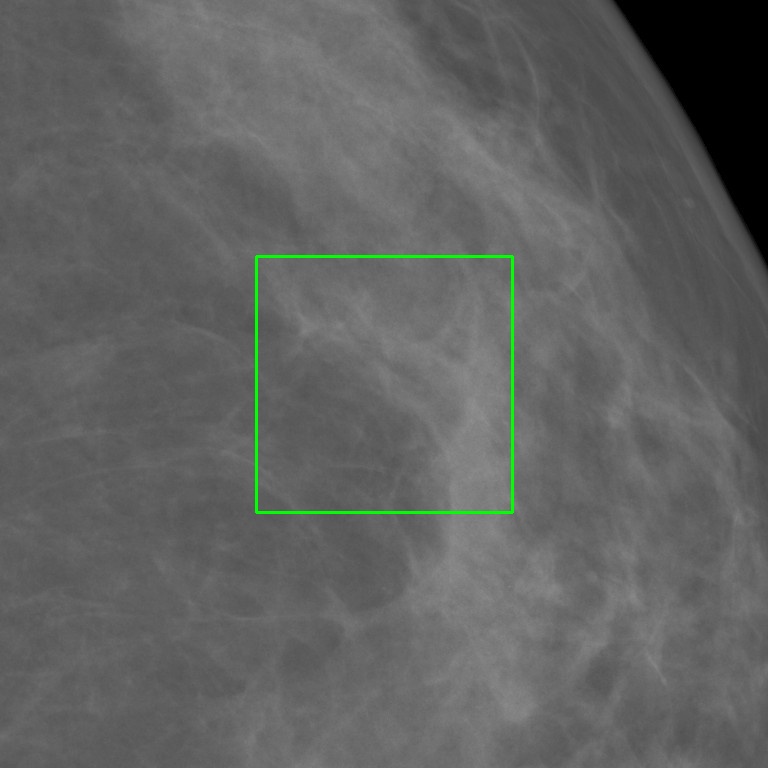}
    \end{subfigure}\hspace{0.1cm}%
    \begin{subfigure}[b]{0.22\textwidth}
        \includegraphics[width=\textwidth]{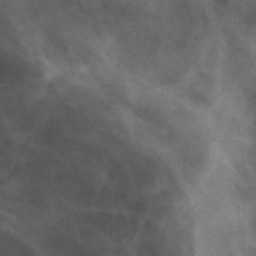}
    \end{subfigure}\hspace{0.1cm}%
    \begin{subfigure}[b]{0.22\textwidth}
        \includegraphics[width=\textwidth]{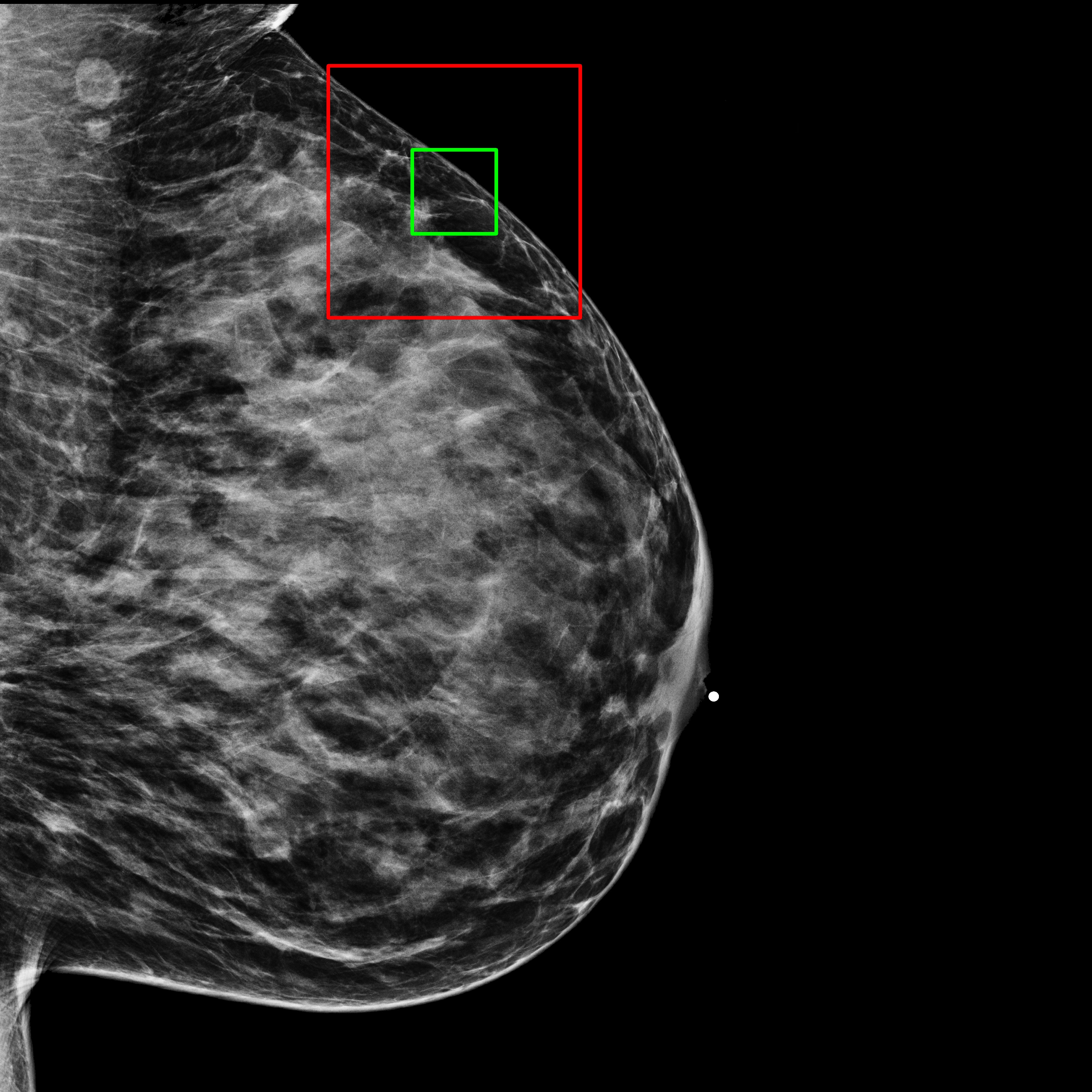}
        \caption{Global context.}
    \end{subfigure}\hspace{0.1cm}%
    \begin{subfigure}[b]{0.22\textwidth}
        \includegraphics[width=\textwidth]{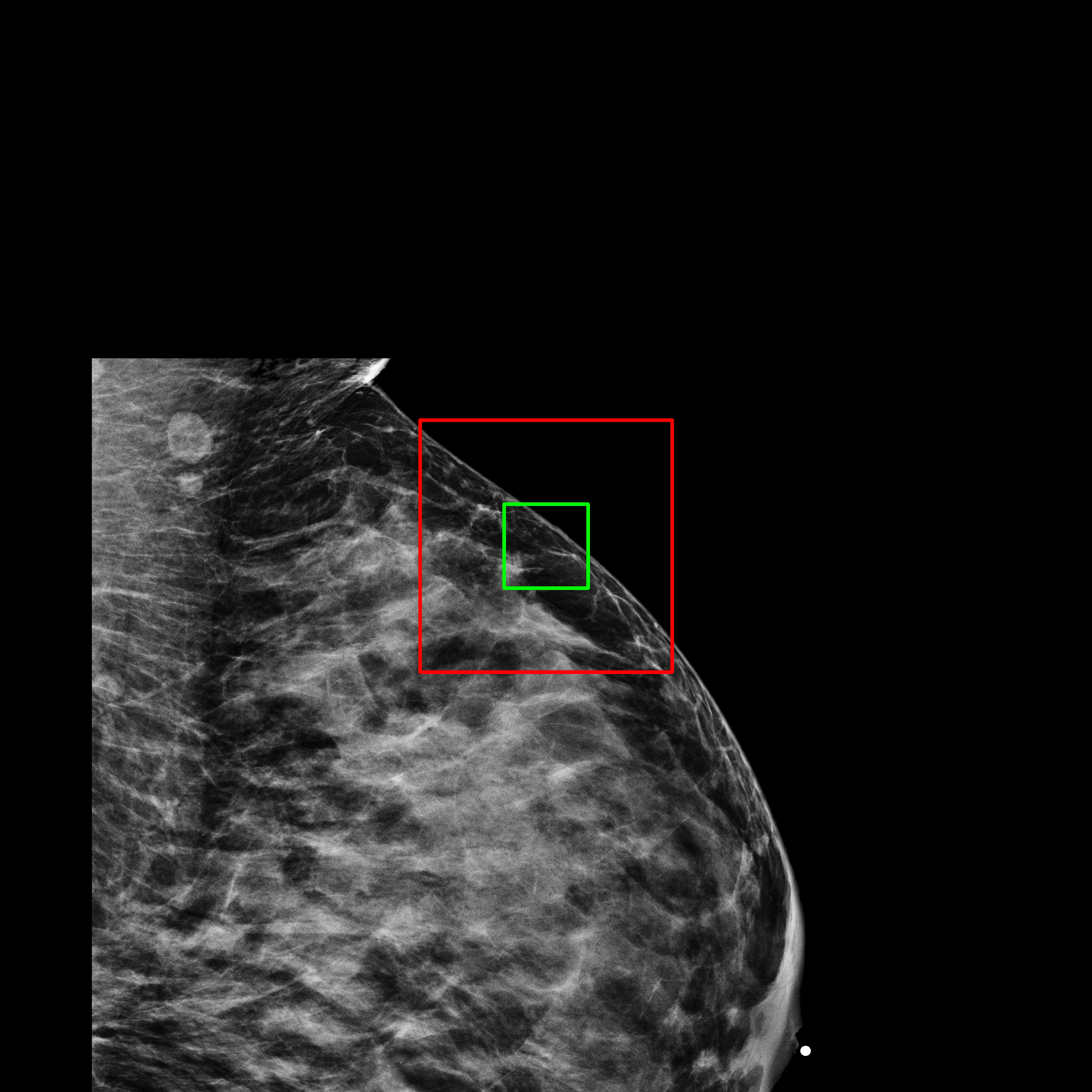}
        \caption{Shifted global context.}
    \end{subfigure}\hspace{0.1cm}%
    \begin{subfigure}[b]{0.22\textwidth}
        \includegraphics[width=\textwidth]{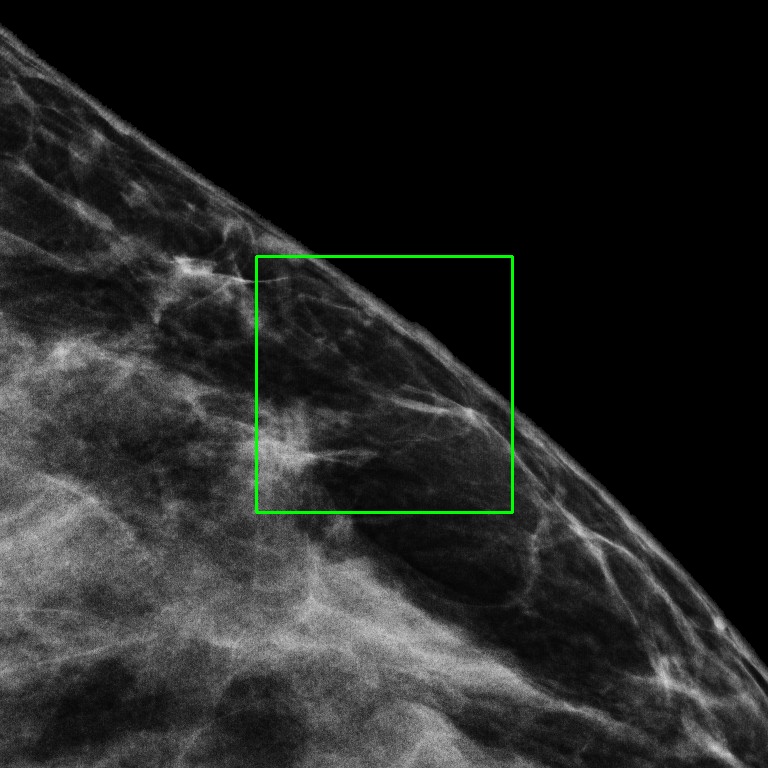}
        \caption{Local context.}
    \end{subfigure}\hspace{0.1cm}%
    \begin{subfigure}[b]{0.22\textwidth}
        \includegraphics[width=\textwidth]{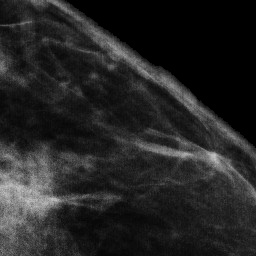}
        \caption{Patch.}
    \end{subfigure}\hspace{0.1cm}%
    \caption{Examples of input per channel. a) Whole image from the dataset scaled down to $256\times256$. (b) Shifted whole image from the dataset scaled down to $256\times256$. (c) Local context, a $768\times768$-pixel patch of the original image, scaled down to $256\times256$ pixels. (d)  Full-resolution $256\times256$ patch of the original image. The mid-resolution model uses (a) and (b) to generate (c). The high-resolution model uses (a) and (c) to generate (d). The specific parts are labeled in the context images, with the patch indicated in green and the local context in red.}
    \label{fig:input}
\end{figure*}

Fig. \ref{fig:input} shows samples of per-channel inputs for the mid-resolution and high-resolution models in the MAMBO pipeline.

\input{images/figures/ablation_figs}

\onecolumn

\section{Examples of High-Resolution Images Generated by MAMBO}
\label{supp:additional_samples}

\begin{figure*}[htp]
\centering
\includegraphics[width=0.495\textwidth]{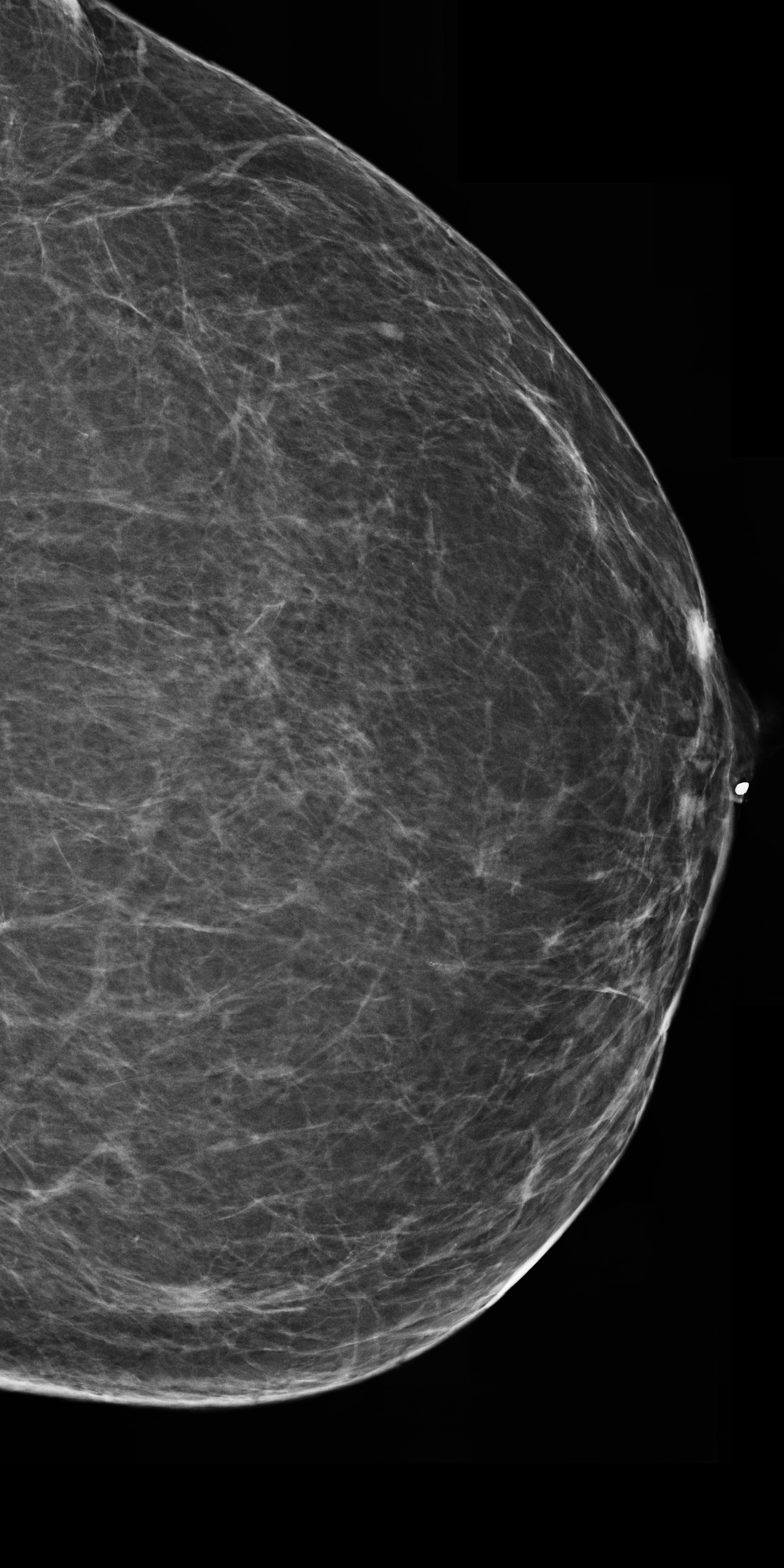}
\includegraphics[width=0.495\textwidth]{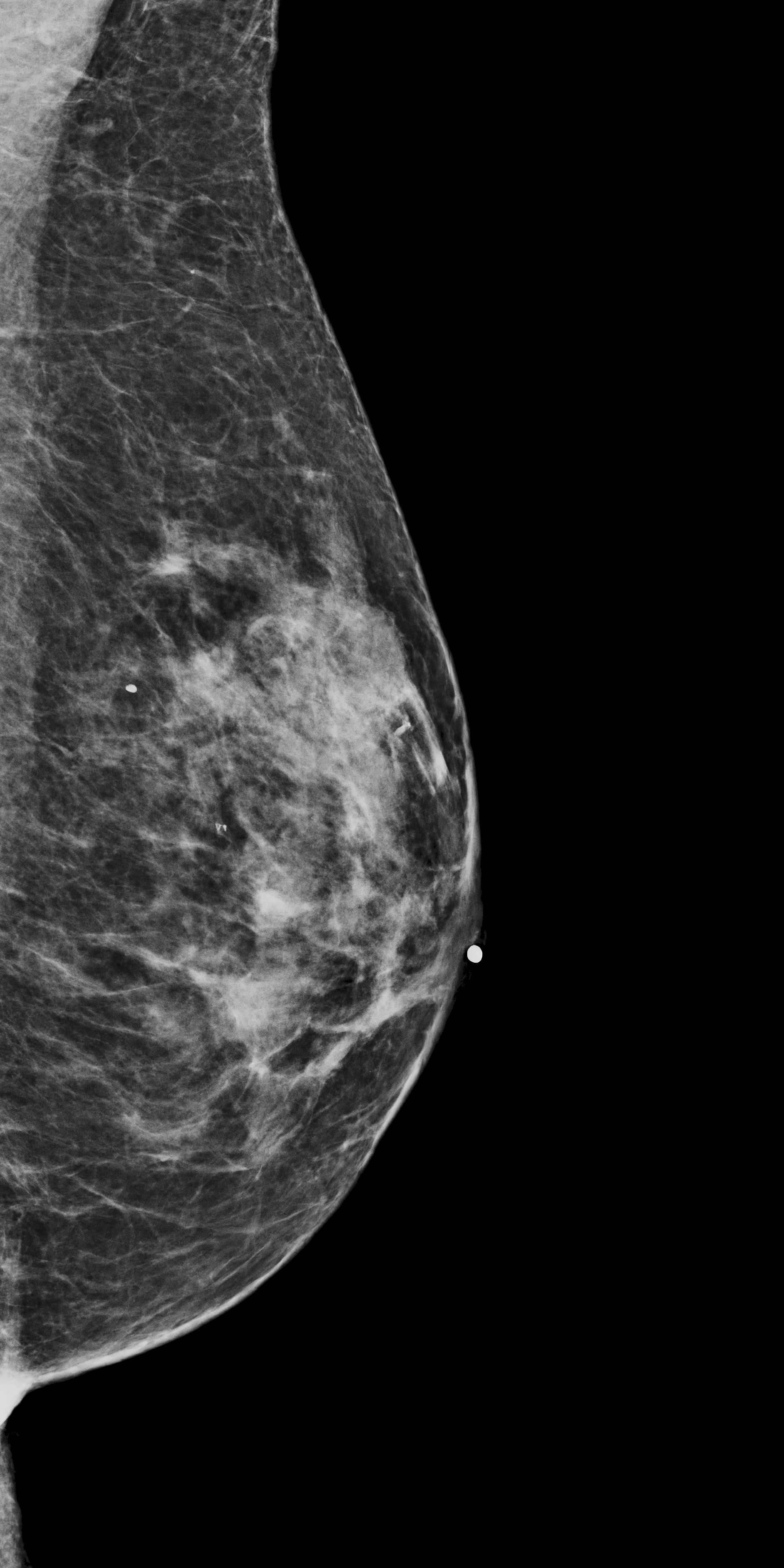}
\caption{
Synthetic full-resolution images ($3840\times3840$ pixels) generated using MAMBO trained on the RSNA dataset.
}
\label{fig:rsna_full_res}
\end{figure*}

\begin{figure*}[htp]
\centering
\includegraphics[width=0.495\textwidth]{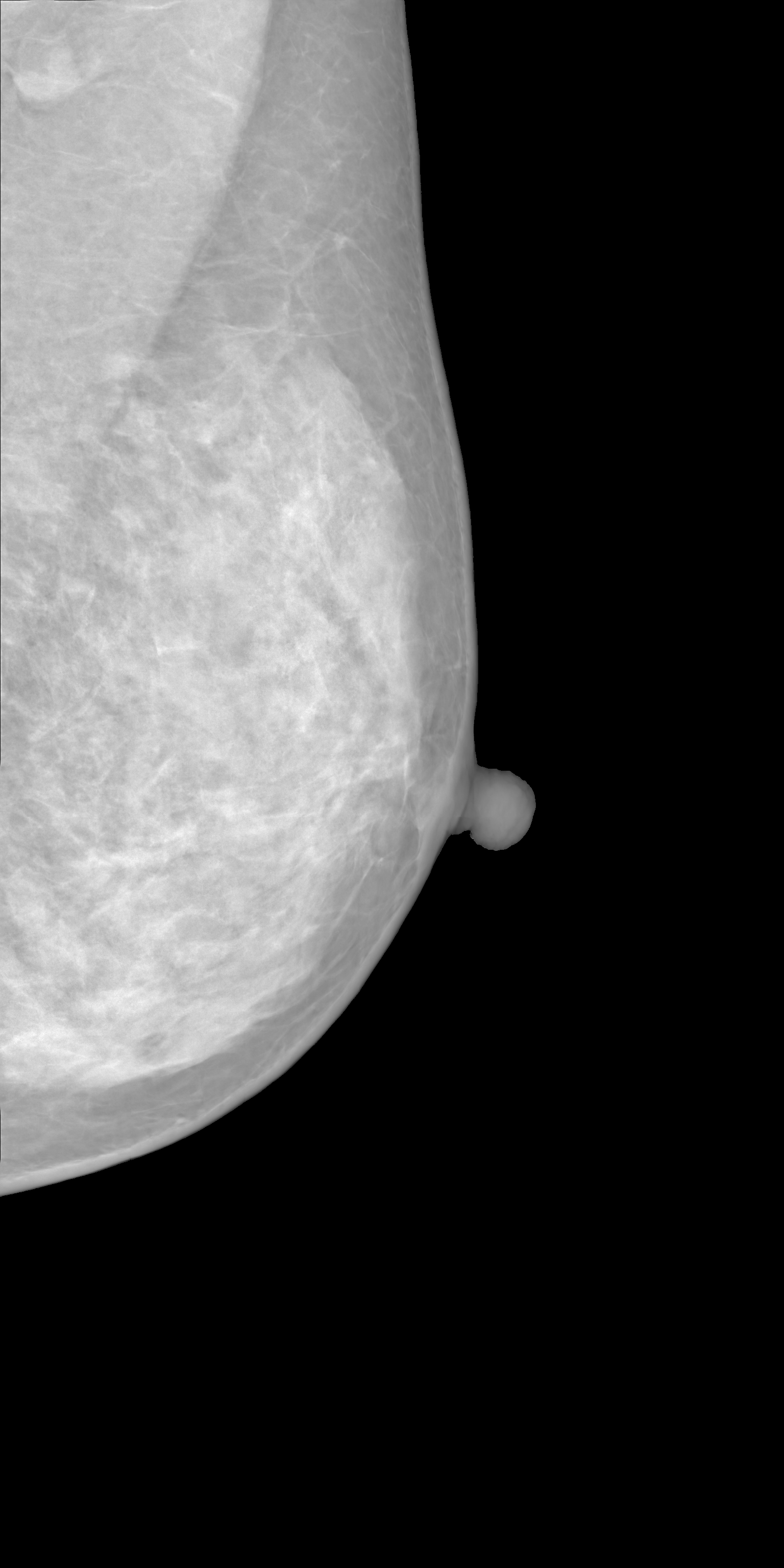}
\includegraphics[width=0.495\textwidth]{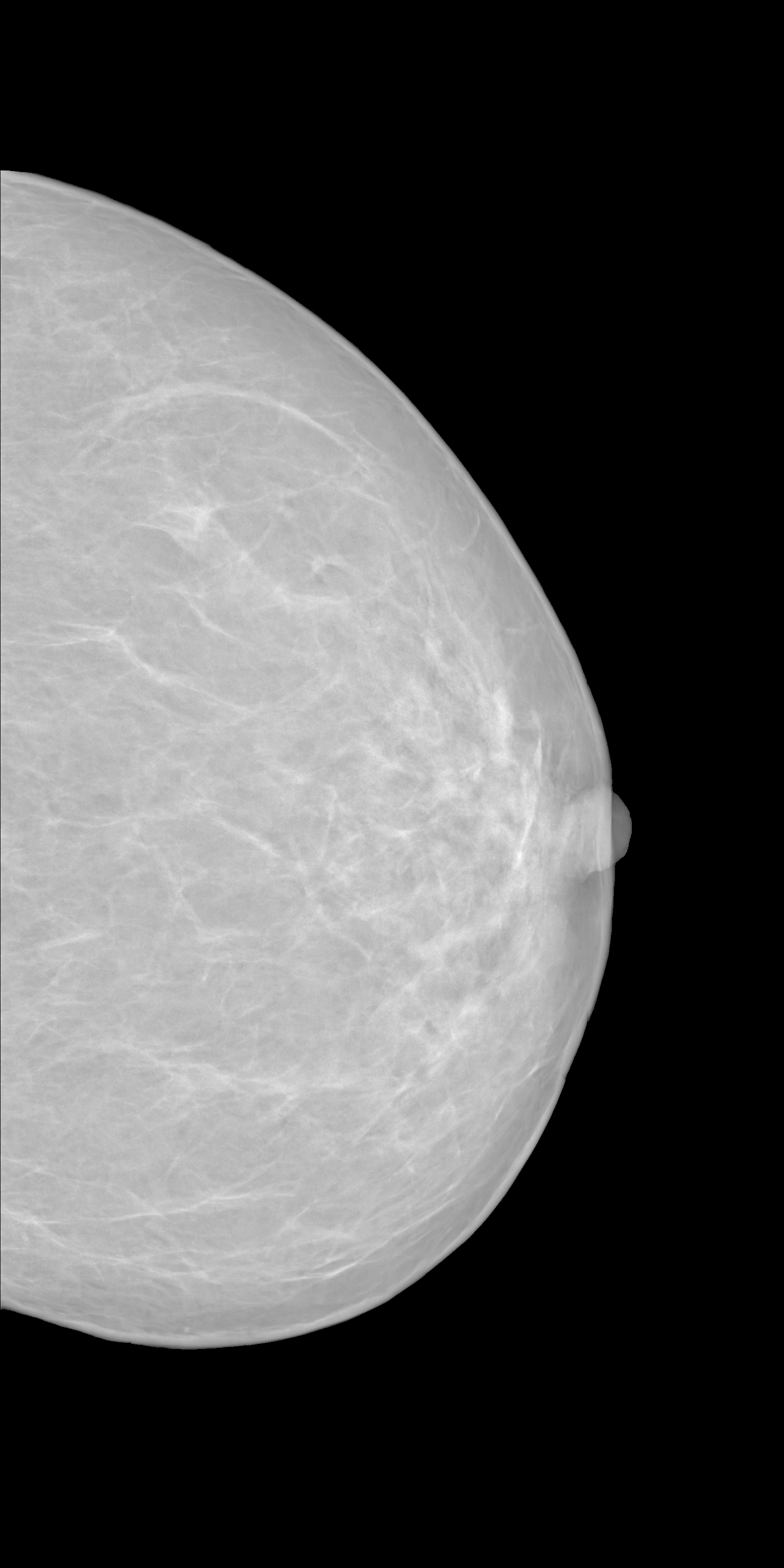}
\caption{
Synthetic full-resolution images ($3072\times3072$ pixels) generated using MAMBO trained on the VinDr dataset.
}
\label{fig:vindr_full_res}
\end{figure*}

The mammograms generated using MAMBO (all three models connected in a pipeline) are shown on Fig. \ref{fig:rsna_full_res} and Fig. \ref{fig:vindr_full_res}. The model is designed to generate full-square images to accommodate varying breast proportions. However, the provided examples are cropped to focus solely on the breast region.

\clearpage
\section{Examples of Lesions on Images Generated by MAMBO Annotated by Experts}
\label{supp:expert_annotation}

\begin{figure*}[htp]
\centering
\includegraphics[width=0.495\textwidth]{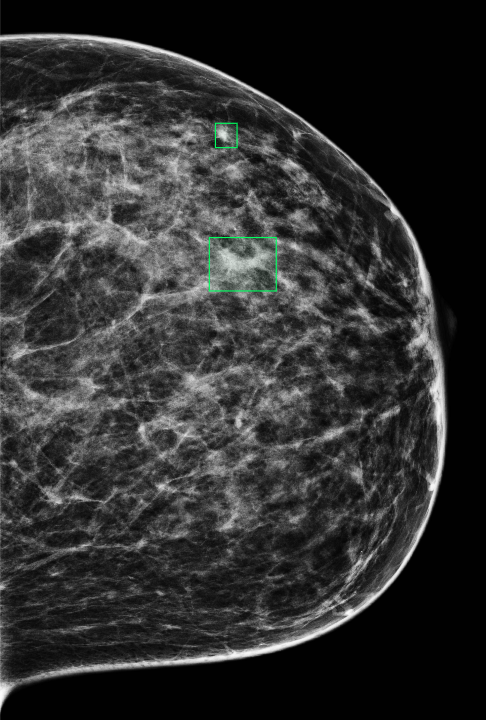}
\includegraphics[width=0.495\textwidth]{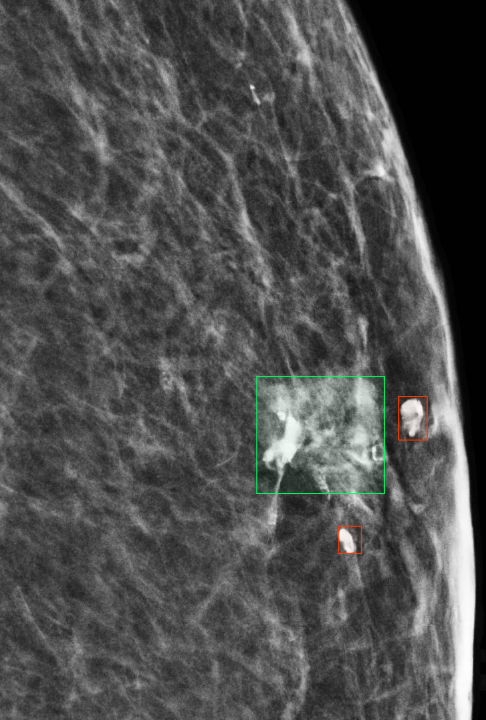}
\caption{
Expert radiologist annotations on masses (green) and calcifications (red).
}
\label{fig:annotations}
\end{figure*}

Realistically generated clinically relevant features (such as masses and calcifications) spotted and annotated by expert radiologists are shown on Fig. \ref{fig:annotations}.

%% file: tables/model_hyperparameters.tex
\begin{table}[ht]
\centering
\caption{MAMBO models architectures}
\label{tab:hyperparameters}
{\resizebox{\columnwidth}{!}{
\begin{tabular}{lccc}
\hline
\textbf{Hyperparameter}     & \textbf{Stage 1} & \textbf{Stage 2} & \textbf{Stage 3} \\
\hline
Patch input size   & $1\times256\times256$       & $3\times256\times256$       & $3\times256\times256$ \\
Initial number of channels            & 128     & 128      & 128      \\
Channel multiplier          & [1, 2, 2, 4, 4] & [1, 2, 2, 4, 4]         & [1, 2, 2, 4, 4]    \\
Learning rate               & $5\times 10^{-5}$ & $5\times 10^{-5}$     & $5\times 10^{-5}$   \\
Batch size                  & 8               & 8               & 8                \\
Number of diffusion steps   & 1000             & 1000             & 1000             \\
Noise scheduler             & Linear           & Linear           & Linear           \\
Optimizer                   & Adam             & Adam             & Adam             \\
Number of parameters        & 78.24M              & 78.25M            & 78.25M             \\
\hline
\end{tabular}
}}
\end{table}

%% file: tables/lambda.tex
\begin{table}[htp]
\centering
\small

{\resizebox{\columnwidth}{!}{
\begin{tabular}{lccccccc}
\toprule
\textbf{$\lambda$} & 300 & 400 & 500 & 600 & \textbf{700} & 800 & 900 \\
\midrule
\textbf{IoU} $\uparrow$ & 0.082 & 0.139 & 0.157 & 0.198 & \textbf{0.216} & 0.200 & 0.207 \\
\bottomrule
\end{tabular}
}}
\caption{Average IoU for different timesteps $\lambda$ over the whole dataset.}
\label{tab:buckets}
\end{table}

%% file: images/figures/ablation_figs.tex
\begin{figure*}[htbp]
    \centering
    \begin{subfigure}[b]{0.22\textwidth}
        \includegraphics[width=\textwidth]{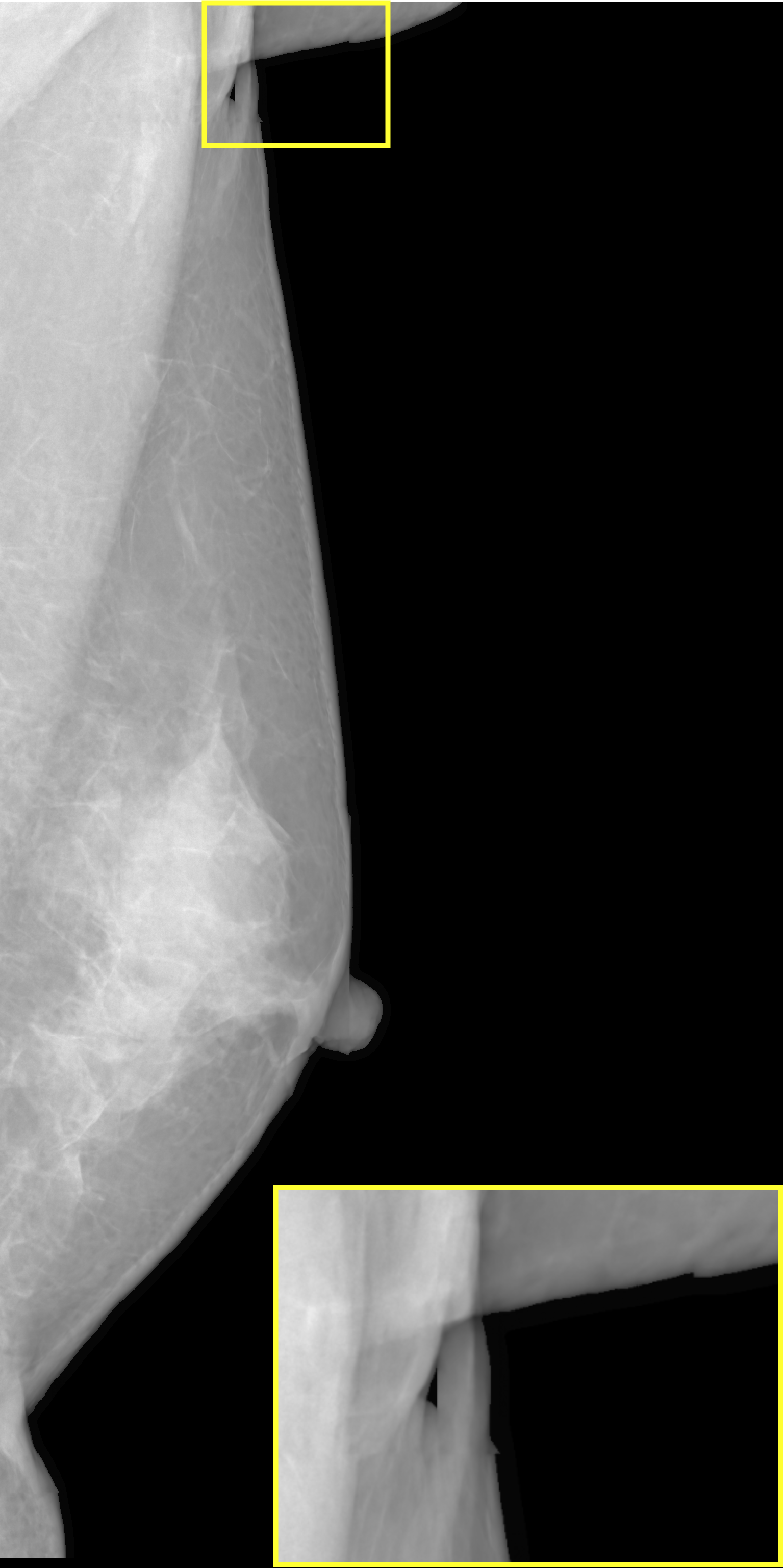}
        \label{fig:lcl_256_1}
        \caption{}
    \end{subfigure}
    \begin{subfigure}[b]{0.22\textwidth}
        \includegraphics[width=\textwidth]{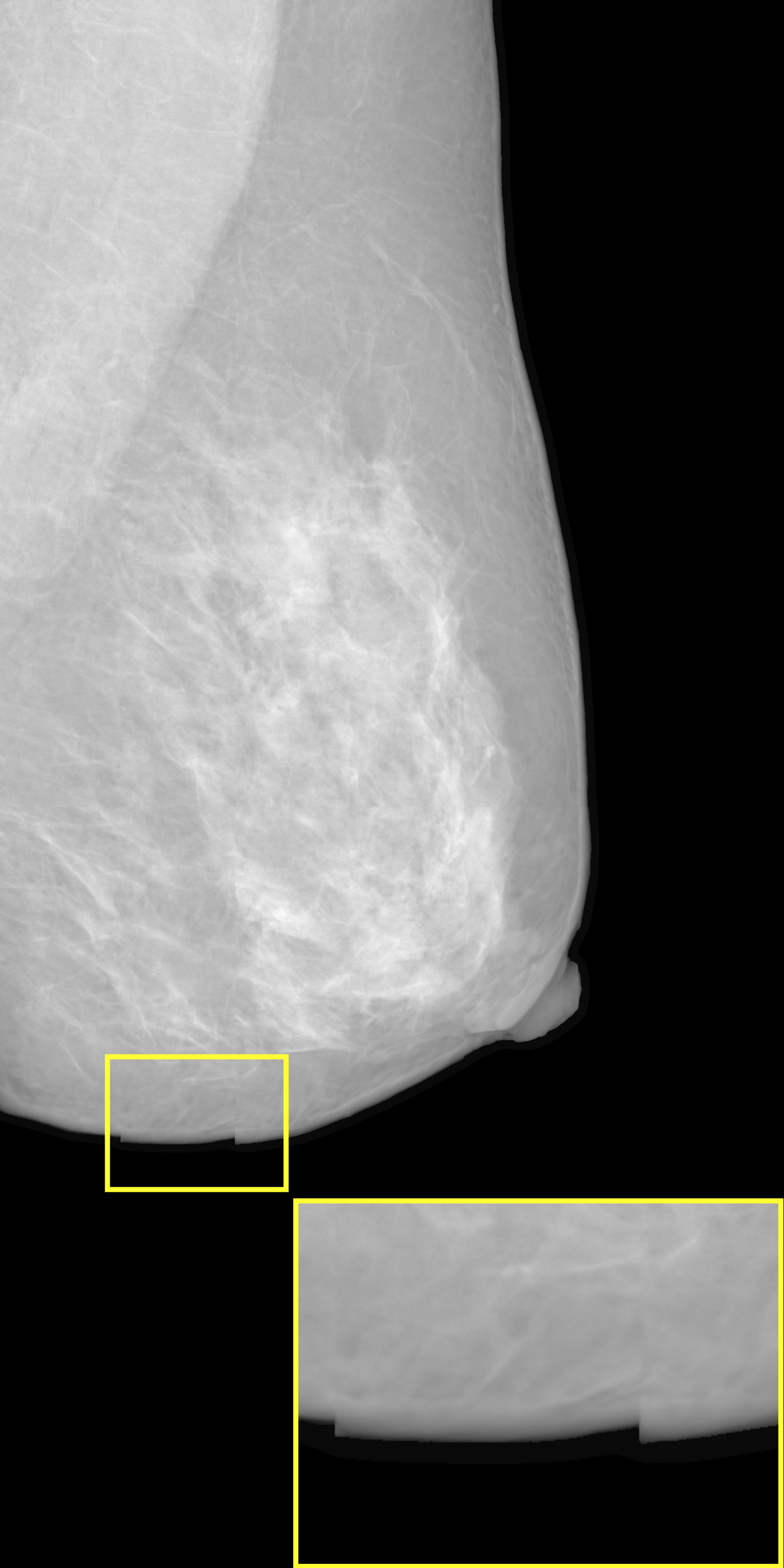}
        \label{fig:lcl_256_2}
        \caption{}
    \end{subfigure}
\begin{subfigure}[b]{0.22\textwidth}
        \includegraphics[width=\textwidth]{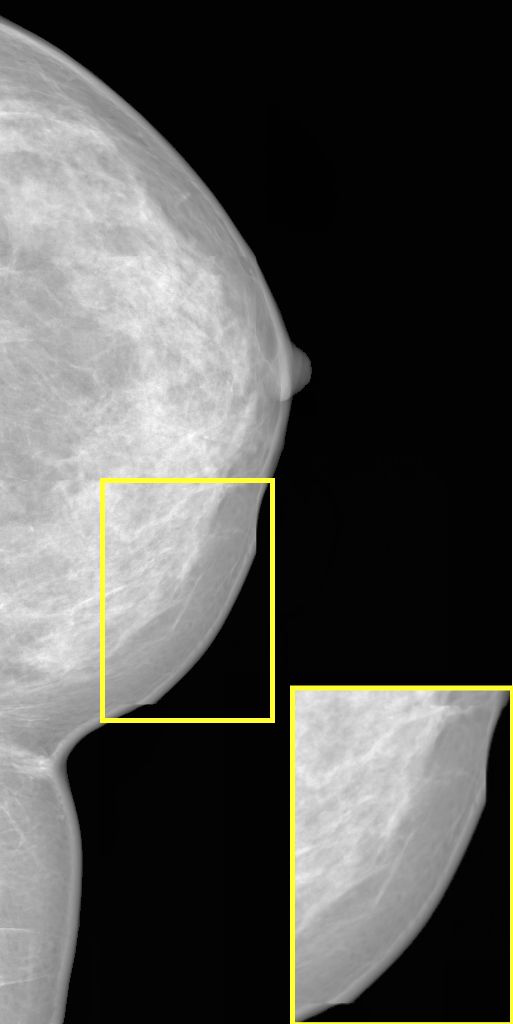}
        \label{fig:model_2c_1}
        \caption{}
    \end{subfigure}
    \begin{subfigure}[b]{0.22\textwidth}
        \includegraphics[width=\textwidth]{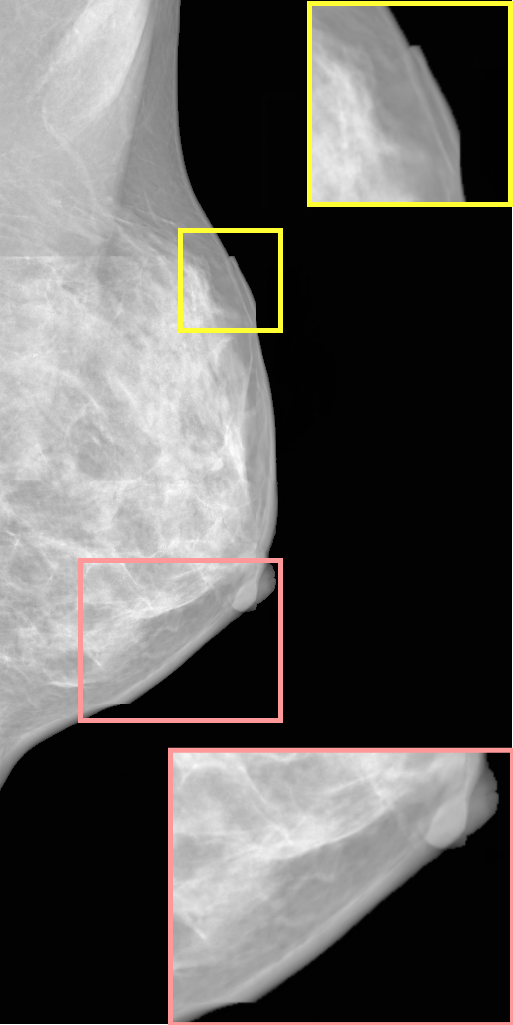}
        \label{fig:model_2c_2}
        \caption{}
    \end{subfigure}
    \caption{(a, b) Visible artifacts at the full-resolution images ($3072\times 3072$) generated in one stage, using model Local context 256.
    (c, d) Visible artifacts at the middle-sized images ($1022 \times 1022$) generated using MAMBO w/o gl. ctx.}
    \label{fig:ablation_bad}
\end{figure*}